\documentclass[final,times,3p,sort&compress,comma]{elsarticle}

\usepackage{lineno,hyperref}
\usepackage{color}
\journal{Prog. Surf. Sci.}

\newcommand{\beql}[1]{\begin{equation}\label{#1}}
\newcommand{\beq}{\begin{equation}}
\newcommand{\eeq}{\end{equation}}
\newcommand{\ben}{\begin{eqnarray}}
\newcommand{\een}{\end{eqnarray}}

\newcommand{\av}[1]{\langle #1 \rangle}

\newcommand{\revision}[1]{{\color{black}{#1}}}
\begin{document}

\begin{frontmatter}

\title{Semi-classical generalized Langevin equation for equilibrium and nonequilibrium molecular dynamics simulation}
%% Group authors per affiliation:
\author[add1]{Jing-Tao L\"u}
\ead{jtlu@hust.edu.cn}
\author[add1]{Bing-Zhong Hu}
\author[add2]{Per Hedeg{\aa}rd}
\author[add3]{Mads Brandbyge}
\address[add1]{School of Physics and Wuhan National High Magnetic Field Center, Huazhong University of Science and Technology, 430074 Wuhan, P. R. China}
\address[add2]{Niels-Bohr Institute \& Nano-science Center, University of Copenhagen, 2100 Copenhagen \O, Denmark}
\address[add3]{Department of Micro- and Nanotechnology, Technical University of Denmark, DK-2800 Kongens Lyngby, Denmark}

\begin{abstract}
Molecular dynamics (MD) simulation based on Langevin equation has been widely used
in the study of structural, thermal properties of matters in difference phases.
Normally, the atomic dynamics are described by classical equations of motion
and the effect of the environment is taken into account through the fluctuating
and frictional forces. Generally, the nuclear
quantum effects and their coupling to other degrees of freedom are difficult to
include in an efficient way. This could be a serious limitation on its
application to the study of dynamical properties of materials made from light
elements, in the presence of external driving electrical or thermal fields. One
example of such system is single molecular dynamics on metal surface, an
important system that has received intense study in surface science. In this
review, we summarize recent effort in extending the Langevin MD
to include nuclear quantum effect and their coupling to flowing electrical
current. We discuss its applications in the study of adsorbate dynamics 
on metal surface, current-induced dynamics in molecular junctions,
and quantum thermal transport between different reservoirs.
\end{abstract}


\end{frontmatter}

%\linenumbers

%%%%%%%%%%%%%%%%%%%%%%
\section{Introduction}
%%%%%%%%%%%%%%%%%%%%%%
The Langevin equation has
been widely used to describe the dynamics of open systems interacting with an
environment (bath). Their interaction introduces dissipation and fluctuations
to the system\cite{senitzky_dissipation_1960,senitzky_dissipation_1961,lax_formal_1963}, 
which are incorporated into the Langevin equation as friction and noise terms. 
When the time scale of the particle is comparable to that
of the environmental degrees of freedom (DoF), the frictional force felt by the
particle will have a memory kernel, meaning that the friction acting on the
particle depends on the velocity at an earlier time. This leads to the generalized 
Langevin equation (GLE)\cite{mori_transport_1965,nordholm_systematic_1975,ingold_path_2002,grabert_quantum_1984,hanggi_fundamental_2005}. 
By solving the GLE, different equilibrium and
nonequilibrium mechanical, thermal properties of the system can be extracted.  

Although the studied system could be made from different kinds of DoF\cite{Gardiner}, the most widely studied one is nuclear or atomic 
or phononic DoF under the influence of thermal baths\cite{adelman_generalized_1974,adelman_generalized_1976,dhar_heat_2001,segal_thermal_2003,wang_quantum_2007,kantorovich_generalized_2008-1,kantorovich_generalized_2008,ceriotti_nuclear_2009,dammak_quantum_2009}.
Interesting applications include the study of nuclear quantum effects\cite{ceriotti_nuclear_2009,dammak_quantum_2009,dammak_isotope_2012,ceriotti_colored-noise_2010,brieuc_zero-point_2016,brieuc_quantum_2016,ganeshan_simulation_2013,bronstein_quantum_2016,bronstein_quantum-driven_2014,guo_atomic-scale_2017}, heat transport between two different thermal
baths\cite{dhar_heat_2001,dhar_quantum_2003,segal_thermal_2003,lee_heat_2005,dhar_heat_2006,wang_quantum_2007,wang_quantum_2008,roy_crossover_2008,kantorovich_generalized_2008-1,kantorovich_generalized_2008,lu_coupled_2009,wang_molecular_2009,stella_generalized_2014,ness_applications_2015,li_colloquium_2012,kosevich_effects_2013}, scattering of single molecule on surfaces\cite{nieto_reactive_2006,juaristi_role_2008,jiang_electronhole_2016,novko_ab_2015,fuchsel_enigmatic_2016,grotemeyer_electronic_2014,xin_strong_2015,fuchsel_dissipative_2011,bunermann_electron-hole_2015,blanco-rey_electronic_2014,kruger_no_2015,rittmeyer_electronic_2015,askerka_role_2016,maurer_mode_2017,galperin_nuclear_2015,dou_frictional_2015,dou_universality_2017,dou_electronic_2017,dou_born-oppenheimer_2017},
 and so on. 
 %\mbnote{We should put the corresponding citations when we list different things.}

Classical GLEs derived from Newtonian equation of motion can also be extended to the
quantum mechanical regime, using the Heisenberg equation of
motion\cite{ford_statistical_1965}, the influence functional
approach of Feynman \& Vernon\cite{feynman_theory_1963}, and the density 
matrix method\cite{mccaul_partition-free_2017,kantorovich_c_2016}. Caldeira and Leggett
successfully used the influence functional approach to study quantum tunneling
in macroscopic systems and dynamics of quantum Brownian
motion\cite{caldeira_influence_1981,caldeira_path_1983}. In these studies, the
environment is modeled by an infinite set of harmonic oscillators occupied by
the quantum mechanical Bose-Einstein distribution with the zero-point
fluctuations included. The bilinear coupling of the system to the quantum reservoir introduces partial
quantum mechanical effects to the system, even if the system itself follows
the classical equations of motion\cite{caldeira_influence_1981,caldeira_path_1983,schmid_quasiclassical_1982,prezhdo_quantized_2000,prezhdo_quantized_2002,prezhdo_non-adiabatic_2003,prezhdo_quantized_2006}. This semi-classical
GLE (SGLE)  has been used recently to study the nuclear quantum effect
\cite{dammak_quantum_2009,ceriotti_langevin_2009,ceriotti_nuclear_2009,ceriotti_efficient_2012,ceriotti_nuclear_2016}.
If the system couples to reservoirs with different temperatures,
it can also be used to study the dynamics of heat
transport \cite{dhar_heat_2001,dhar_heat_2006,wang_quantum_2007,wang_molecular_2009}.

The extension of the influence functional approach to consider the electronic 
reservoir was also conducted by several researchers and
compared to harmonic oscillator
reservoir\cite{guinea_friction_1984,chang_dissipative_1985,hedegard_quantum_1987,hedegrd_light_1987}.
It has been used to study muon diffusion in
metals, single molecule scattering, vibrational relaxation on metal surface, and so on. 
The electron-hole pair (EHP) excitation is the origin of the
friction force felt by the system, termed electronic friction\cite{persson_electronic_1982,hellsing_electronic_1984,avouris_excited_1984,persson_surface_1991,headgordon_vibrational_1992,head-gordon_molecular-orbital_1992,headgordon_molecular_1995}. In surface
science, molecular dynamics (MD) incorporating electronic friction has proven useful in the study of
adsorbate dynamics on metal surface, where the metal electrons couple to the
atomic DoF and damp their motion\cite{headgordon_molecular_1995}. 

In the important new case of a nonequilibrium electron environment, i.e., in the presence of electrical
current, the SGLE can also be used to study current-induced
forces and Joule heating in molecular conductors and nanomechanical systems\cite{brandbyge_theory_1994,brandbyge_electronically_1995,mozyrsky_quantum-limited_2004,mozyrsky_intermittent_2006,daligault_non-adiabatic_2017,lu_blowing_2010,lu_laserlike_2011,lu_current-induced_2011,lu_current-induced_2012,gunst_phonon_2013,lu_current-induced_2015,bode_scattering_2011,bode_current-induced_2012,hussein_semiclassical_2010,metelmann_adiabaticity_2011,metelmann_transport_2012,lopez-monis_limit_2012,mosshammer_semiclassical_2014}.

The scope of present review is to summarize recent advances and applications of
the SGLE to model MD in contact with electronic and phononic reservoirs
possibly in nonequilibrium situations. 
First, we will briefly sketch the derivation of the SGLE from the
influence functional approach, taking an electronic reservoir as an example. Within the
harmonic approximation, we will
analyze the effect of the non-thermal, nonequilibrium electronic environment 
on the nuclear dynamics. 
This is followed by several applications in the study of thermal transport, nuclear
quantum effects, and current-induced dynamics. Finally, we give a brief summary and 
perspective for future developments.

%%%%%%%%%%%%%%%%%%%%%%
\section{Theory}
%%%%%%%%%%%%%%%%%%%%%%
Our starting point is the separation of the whole world into system and environment.
Here the system is the atomic DoF that we are interested in, and
the environment is the rest of the world. Our goal is to derive an equation of
motion for the system DoF. The first step is to write down
equations of motion for the system plus the environment DoF.  We
then eliminate the environment degrees of freedom and obtain effective equations of motion of
the system including the effects of the environment, yielding dissipation and 
fluctuation terms. 

Different approaches can be used to perform this procedure\cite{feynman_theory_1963,ford_statistical_1965,caldeira_influence_1981,caldeira_path_1983,guinea_friction_1984,chang_dissipative_1985,hashitsume_derivation_1986,hedegard_quantum_1987,gardiner_quantum_1988,adelman_generalized_1974,adelman_generalized_1976,dhar_heat_2001,dhar_quantum_2003,segal_thermal_2003,lee_heat_2005,dhar_heat_2006,wang_quantum_2007,kantorovich_generalized_2008,kantorovich_generalized_2008-1,kantorovich_c_2016,mccaul_partition-free_2017,dou_frictional_2015,kleinert_quantum_1995,ingold_path_2002,hanggi_fundamental_2005}.
Our choice is the influence functional approach of Feynman and
Vernon\cite{feynman_theory_1963}.  In this approach, one starts with the full
density matrix including both electrons and nuclei. Selecting the system as
the nuclear DoF that one is interested in, one tries to obtain the reduced density
matrix of the system only. This is realized by tracing out the environment DoF.
The influence functional describes the effect of the environment on the system.
From the reduced density matrix, one then performs an expansion over the
classical nuclear paths to the second order, taking the deviation from the
classical path as perturbation.  It has at least two advantages: (1) it can deal
with both boson and fermion reservoirs; (2) the reservoirs may be in a
nonequilibrium state due to external driving\cite{hanggi_fundamental_2005}.  
A system coupling to a bath of
harmonic oscillators has been considered in seminal works by
Feynman and Vernon\cite{feynman_theory_1963},
Caldeira and Leggett\cite{caldeira_influence_1981,caldeira_path_1983}.  We here take
a noninteracting electronic reservoir as an example. 

%+++++++++++++++++++++++++++++++
\subsection{Influence functional}
%+++++++++++++++++++++++++++++++
We only give a sketch of the derivation and the details can be found in our earlier
works\cite{lu_blowing_2010,lu_current-induced_2012}.  
We consider a system including electrons and nuclei. The total Hamiltonian has two parts
\begin{equation}
H = H_e(x) + H_I.
\end{equation}
The nuclear Hamiltonian takes the standard form, including the kinetic and the potential energy terms
\begin{equation}
H_I = \sum_i \frac{p_i^2}{2M_i} + V_I(x).
\end{equation}
The electrons couple to the nuclear DoF through a displacement dependent Hamiltonian $H_e(x)$, 
where $x$ represents a vector made from displacement of the nuclear DoF
\begin{equation}
	H_e(x) = \int dr\; \Psi(r)^\dagger \left[ H_0+H_{eI}(x) \right] \Psi(r).
	\label{}
\end{equation}
Here, $H_0$ is purely electronic, the coupling to nuclear DoF is in $H_{eI}(x)$, through it's dependence on $x$. The operators $\Psi(r)$ and $\Psi(r)^\dagger$ are the creation and annihilation field operator of
electrons, and $r$ represents the electron position. 
\revision{Here, we did not include the electron-electron interaction explicitly. 
We assume that $H_e$ represents either a single electron Hamiltonian, or includes electron-electron
interaction at a mean field level, like by density functional theory (DFT).
}

The reduced density matrix of the system in the displacement
representation $\rho_s(x,y,t)$ can be written as
\begin{eqnarray}
	\rho_s(x_2,y_2,t_2) &=& \int dx_1 \int dy_1 \; \mathcal{K}(x_2,y_2,t_2;x_1,y_1,t_1)\rho_s(x_1,y_1,t_1),
\end{eqnarray}
with the propagator of the reduced density matrix being (we use $\hbar=1$ in all formulas throughout the paper)
\begin{equation}
	\mathcal{K} (x_2,y_2,t_2;x_1,y_1,t_1) = \int_{x_1}^{x_2}\mathcal{D}x \int_{y_1}^{y_2}\;  \mathcal{D}y e^{i[S_s(x)-S_s(y)]} F(x,y).
\end{equation}
Here, $x$ and $y$ are a pair of displacement histories of the nuclei from time
$t_1$ to time $t_2$,  $(x_1,x_2)$ and $(y_1,y_2)$ correspond to the forward and
backward propagating paths of the reduced density matrix,  and  $S_{s}(x)$ and $S_{s}(y)$  are the system
action along these two paths, respectively.  The influence functional $F(x,y)$
includes the information of coupling to the electronic environment via the electronic time-propagators on forward and backward paths, $U(t,x)$ an $U(t,y)$, respectively,
\begin{eqnarray}
	%F[x,y]=\frac{{{\rm Tr}_e[\rho_e U^\dagger(t,y)U(t,x)]}}{{\rm Tr}_e[\rho_e ]}.
	F(x,y)={{\rm Tr}_e[\rho_e U^\dagger(t,y)U(t,x)]}/{\rm Tr}_e[\rho_e ].
\end{eqnarray}
Here, $\rho_e$ is the initial density matrix of the electron reservoir when the nuclei are at
their initial positions. 

At this stage, it is convenient to introduce two new variables, which are the
average and difference of the forward $x$ and backward $y$ displacement,
respectively
\begin{equation}
R(t) = \frac{1}{2}[x(t)+y(t)],\quad \xi(t) = x(t)-y(t).
\end{equation}
The average path $R(t)$ describes the propagation of the diagonal matrix
element of the reduced nuclear density matrix $\rho_s(R+\xi/2,R-\xi/2)$.  Its
propagator can be written in terms of the new variables
\begin{equation}
	\mathcal{K}(R_2,\xi_2,t_2;R_1,\xi_1,t_1)  = \int_{R_1}^{R_2}\mathcal{D}R \int_{\xi_1}^{\xi_2} \mathcal{D}\xi  e^{i\left[S_s(R+\xi/2)-S_s(R-\xi/2)\right]} F\left(R+\xi/2,R-\xi/2\right).
\end{equation}
The action of the nuclear part is
\begin{eqnarray}
{S_s(R+\xi/2)-S_s(R-\xi/2)} 
&=& \int_{t_1}^{t_2} dt' \left[M \dot{R}\,\dot{\xi} - V_I(R+\xi/2)+V_I(R-\xi/2)\right]\nonumber\\
&\approx&-\int_{t_1}^{t_2} dt' \left[M \ddot{R} + \frac{\partial V_I(R)}{\partial R}\right] \xi.
\end{eqnarray}Here, $\dot{R}=\partial R/\partial t'$ is derivative with respect to time $t'$, and we have omitted the time arguments.
In the second equation, we have performed an integration-by-part. The boundary
terms are ignored when performing the integration-by-part since they merely contribute to 
a normalization factor to $\mathcal{K}$. It can be shown that the term
quadratic in $\xi$ is zero.  Thus, to the second order in $\xi$, we get the
classical nuclear equation of motion if we perform the integral over $\xi$.
This shows that, $R(t)$ is actually the classical path that the nuclei would
follow, and $\xi$ measures the fluctuations away from the classical path.

The key problem next is to evaluate $F(x,y)$, and write it as an expansion in $\xi$.  Here, we consider two models of the
electron-nuclear interaction. One is the {\em adiabatic} approximation where the ionic 
velocity $\dot{x}$ is the small
parameter, utilizing the fact that nuclear mass is much larger than that of
electrons. In this case we can perform an expansion over $\dot{x}$ (small-$\dot{x}$
expansion).  In so doing, we may deal with large displacements or even
diffusion of the nuclei.  The resulting Langevin equation becomes Markovian due
to the time scale separation of electronic and nuclear DoF.  The other approach is to
take the displacement itself, $x$, to be small. In this case, the nuclei oscillate in a
small region around their equilibrium positions and we can do a perturbation
expansion over $x$ (small-$x$ expansion). The timescale of nuclei does not
have to be much smaller than that of electrons. Thus, it results in a
generalized Langevin equation with memory kernel.

%++++++++++++++++++++++++++++++++
\subsection{Adiabatic expansion}
%++++++++++++++++++++++++++++++++
Following the standard Fermionic path integral approach, we can write the terms 
in $F(x,y)$ as functional integral of the electronic Grassmann fields $\psi$ and $\psi^*$,
\begin{eqnarray}
	F(x,y)&=&{\rm Tr}_e\left[ \rho_e U^\dagger(t,y) U(t,x) \right] = \int \mathcal{D}(\psi^*\psi)\\
	&\times&{\rm exp}\left[i\int_\gamma d\tau \int dr \psi^*(r,\tau) \left( i\frac{\partial}{\partial \tau}-H_e(\tau) \right) \psi(r,\tau) \right]\nonumber.
	\label{}
\end{eqnarray}
We have combined the forward and backward propagation into one contour with
time $\tau$, $x$ and $y$ are then paths on the upper and lower branches of the
contour.  The electron fields satisfy the boundary condition $\psi(r,t_1) =
-\psi(r,t_2)$.  

In the adiabatic (small-$\dot{x}$) approximation, assuming the nuclear dynamics is much slower than that of the electrons in the
environment, we can perform an expansion over the Born-Oppenheimer eigenfunctions $\phi_n(\tau)$ corresponding to $H_e(t)$ 
at time $t$
\begin{equation}
	\psi(x(\tau)) = \sum_n a_n(\tau) \phi_n(x(\tau)).
	\label{}
\end{equation}
The action of the electrons can be written as
\begin{equation}
	S_e = a^* (G_0^{-1}+V) a,
	\label{}
\end{equation}
with $V_{mn}(\tau,\tau') = i\langle \phi_m(\tau)|\dot{\phi}_n(\tau')\rangle$. We are here employing a very condensed notation, where the time dependence and $n$ dependence of $a_n(\tau)$ is suppressed, and two integrals over times and sums over $n$ are understood, much like the Einstein summation notation, known from relativity theory. The Green's functions $G_0$ is diagonal in the $n$ variable and the diagonal elements are given by
\begin{equation}
	\left( i\frac{\partial}{\partial\tau}-\varepsilon_n(\tau) \right) G_{0nn}(\tau,\tau') = \delta(\tau,\tau').
	\label{}
\end{equation}
In this way, the functional integral can be performed formally,
\begin{eqnarray}
	\int \mathcal{D}(a^*a)\; e^{iS_e} &=& {\rm det}\left( G_0^{-1}+V \right)={\rm exp}\left[ {\rm Tr}{\rm ln}\left( G_0^{-1}+V \right) \right] \nonumber\\
	&\approx& {\rm exp}\left[ {\rm Tr}{\rm ln}G_0^{-1} +{\rm Tr}\left( G_0V \right) -\frac{1}{2}{\rm Tr}\left( G_0VG_0V \right)\right].
	\label{eq:if2}
\end{eqnarray}
We have expanded it to the 2nd order in the interaction $V$. Actually, we can rewrite $V$ as,
\begin{equation}
V_{mn}(t',t) = i\langle \phi_m(t')|\dot{\phi}_n(t)\rangle = i\langle \phi_m(t')|\partial_{x_j}{\phi_n(t)}\rangle \dot{x}_j.
\end{equation}
After some manipulation of the influence functional, we can show that: (1) The
first term  in Eq.~(\ref{eq:if2}) contributes with an effective Born-Oppenheimer
potential ($V_e$) to the effective action; (2) The second term contributes to a term $\propto \dot{x}
(\dot{y})$ and linear in $\xi$; 
(3) The third term contributes to a term quadratic in $\xi$.  
The total effective action of system can then be written as,
\begin{eqnarray}
S_{\rm tot} =&-&\int_{t_1}^{t_2} dt' \int_{t_1}^{t_2}dt''\left[\xi(t')\left(M \ddot{R}(t') + \frac{\partial V_I(R(t'))}{\partial R}+ \frac{\partial V_{e}(R(t'))}{\partial R} + \Gamma_e \dot{R}(t')\right)\right.\nonumber\\
&+& \left.\frac{i}{2}\xi(t')\Pi(t',t'')\xi(t'')\right].
\end{eqnarray}
The effect of the electronic environment on the system dynamics can be deduced
from the effective action. The nuclear equation of motion gains two extra
terms related to the coupling to the environment. The first is the Born-Oppenheimer force, and the second is the
electronic friction.  Moreover, quantum fluctuations around the classical path
$R(t)$ shows up in the term second order in $\xi$. With the help of the
Hubbard-Stratonovich transformation, its effect on the equation of motion can
be interpreted as a classical Gaussian noise acting on the system.  The final result is a semi-classical Langevin equation\cite{schmid_quasiclassical_1982},
which describes the stochastic classical system within a quantum
electronic
environment\cite{horsfield_power_2004,abedi_mixed_2014,abedi_exact_2010}
\begin{equation}
	\ddot{X}  =  F_I(X) + F_e(X) - \Gamma_e \dot{X} + \chi_e.
	\label{eq:lang1}
\end{equation}
Here, $X$ represents a vector made from the mass-renormalized displacements
$X=(\cdots, \sqrt{M_i}R_i, \cdots)$, $F_I$ and $F_e$ are the nuclear and
the Born-Oppenheimer force, respectively, and $\Gamma_e$ is the friction matrix.
The fluctuating force $\chi_e$ describes the quantum and thermal fluctuations away from the
classical path $R(t)$ in $S_{\rm tot}$.  Its average is zero, and the
correlation function is $\av{\chi(t)\chi(t')} = \Pi(t,t')$, which can also be
expressed from $G_0$ and $V$\cite{lu_blowing_2010}. In equilibrium, it is described by the fluctuation-dissipation relation: $\langle \chi(\omega)\chi^*(\omega)\rangle_{\rm eq} = \omega \Gamma_e \coth(\omega/2k_BT)$. Thus, we have a colored noise with 
frequency dependence. It reduces to white noise in the high temperature, 
classical limit, $\langle \chi(\omega)\chi^*(\omega)\rangle_{\rm eq} =  2\Gamma_e k_BT$. 

Similar equations have been derived using the scattering theory
approach\cite{bode_scattering_2011,thomas_scattering_2012,bode_scattering_2011}.
The concept of electronic friction $\Gamma_e$ has been widely used to describe
the energy dissipation in the study molecular scattering, diffusion, rotational
and vibrational relaxation on metal surfaces under external stimulation. In these studies,
it is crucial that the displacement of the nuclei could be very large, and
the parameters entering the SGLE may depend on the position of the nuclei.
We will discuss this issue in Sec.~\ref{sec:app}.

%+++++++++++++++++++++++++++++++++++
\subsection{Perturbation expansion}
%+++++++++++++++++++++++++++++++++++
Alternatively, we can perform an expansion over the displacement
$x$. To do that, we consider the linear in $x$ term in $H_{eI}(x)$, and assume
$x$ is small. This results in a linear electron-phonon coupling term
\begin{equation}
	H_{eI} \approx \left.\frac{\partial H_{eI}}{\partial x}\right|_{x_0} x \equiv M^x x.
	\label{}
\end{equation}
An expansion over $x$ to the second order results in a SGLE of the following form in mass-scaled displacement,
\begin{equation}
	\ddot{X}(t) - F_I(X(t)) =  - \int_{-\infty}^t \Pi^r_e(t-t') X(t')dt' + \chi_e(t).
	\label{eq:lang2}
\end{equation}
This equation looks different from Eq.~(\ref{eq:lang1}). Firstly, it has a memory kernel. The
reason is that we do not have the clear time scale separation between the system and 
environment any more.  Secondly, the velocity dependence is absent. Actually, we can do an integration-by-part
over $t'$ to the first term on the right hand side.
This transforms the dependence on $X(t')$ to $\dot{X}(t')$, and introduces an extra
term that renormalizes the potential felt by the nuclei\cite{wang_molecular_2009}.
The renormalization is absent in the adiabatic expansion, since it is included
in the Born-Oppenheimer force $F_e(X)$.
This difference between the $x$ and $\dot{x}$ expansions is well-known as discussed in Ref.~\cite{caldeira_path_1983}.
We will return to the $\chi_e$-correlation function later.

The advantage of the small-$x$ expansion is that, one can make the harmonic approximation for
the nuclear dynamics. Many interesting effects can be identified even within this simple approximation,
and their key features can be more easily analyzed. This is shown in Sec.~\ref{sec:theo}.

%++++++++++++++++++++++++++++++++
\subsection{Including phonon environment}
%++++++++++++++++++++++++++++++++
If the system furthermore couples linearly to a phonon environment (either in $x$ or $\dot{x}$), 
a SGLE of the same form as Eq.~(\ref{eq:lang1}) or (\ref{eq:lang2}) can be obtained. 
Altogether, each term at the right side of the two equations will now include  
terms from electron and phonon, respectively. For example, Eq.~(\ref{eq:lang2}) changes to
\begin{equation}
	\ddot{X}(t)-F_I(X(t))   =  - \int_{-\infty}^t \Pi^r(t-t') X(t')dt' + \chi(t),
	\label{eq:lang4}
\end{equation}
with $\Pi^r = \Pi^r_e+\Pi^r_{ph}$, and $\chi=\chi_e+\chi_{ph}$.
The first term on the right hand side includes both renormalization and
dissipation. The matrix $\Pi^r(t-t')$ is the retarded self-energy due to
system-environment coupling in the nonequilibrium Green's function (NEGF)
theory.  This connection with the NEGF theory is very favorable in terms of
numerical calculation of realistic systems, since the NEGF theory has been widely used in
the study of transport problems.  The phonon part $\Pi^r_{ph}$ can be obtained
exactly if the system-environment coupling is linear.  On the other hand, it is difficult to
obtain the electronic part $\Pi^r_e$ exactly.  The simplest approach is to take
the lowest order term, corresponding to the polarization-like bubble diagram of
the self-energy evaluated using the unperturbed electron Green's function. The
correlation function of the fluctuating force $\chi$ can also be written in
terms of the self-energies in NEGF theory.  In equilibrium, they are related
through the celebrated fluctuation-dissipation theorem.

%++++++++++++++++++++++++++++++++
%\subsection{%Pros and cons}
%++++++++++++++++++++++++++++++++
We have the following comments on the SGLE:
%\begin{itemize}
(1) The SGLE is a powerful tool to study MD
	within electron and phonon environments. Since it is derived from
	\emph{first-principles}, it can be readily used to study realistic
	systems.  Given the system Hamiltonian, the parameters entering the
	equation can be calculated and no fitting parameters are needed.
(2) Since the environment DoF are noninteracting and treated fully quantum
	mechanically, the quantum statistics of the environment is taken into account.
	The environments DoF fulfill the corresponding Fermi-Dirac or
	Bose-Einstein distribution for electrons or phonons, respectively.
	This is important at low temperature.
(3) The environment is not required to be in equilibrium. When including the 
	electronic environment, we can use it to
	perform MD in the presence of electrical current. This is 
	the focus of this review. The SGLE can also be used to study phonon thermal transport
	by introducing two phonon reservoirs at different temperatures. It has 
	been shown for harmonic systems to reproduce the quantum mechanical
	results exactly\cite{wang_nonequilibrium_2007-1}. While at high enough temperature, the system behaves
	classically, and the SGLE gives correct classical results. 
(4) The memory kernel in Eq.~(\ref{eq:lang4}) makes numerical simulation
	quite expensive. Numerical methods have been introduced to eliminate
	the memory kernel by introducing auxiliary
	variables\cite{ceriotti_langevin_2009,lei_data-driven_2016,kantorovich_generalized_2008,kantorovich_generalized_2008-1,kantorovich_c_2016,stella_generalized_2014,ness_applications_2015,ness_nonequilibrium_2016}.

%%%%%%%%%%%%%%%%%%%%%%%%%%%%%%%%%%%%%%%%%%%%%%%%%
\section{Theoretical Analysis: Harmonic modes coupling to electrons}
\label{sec:theo}
%%%%%%%%%%%%%%%%%%%%%%%%%%%%%%%%%%%%%%%%%%%%%%%%%
In this section, staying in the harmonic approximation, we analyze the SGLE in
different circumstances. We show that, it can describe a varieties of
interesting effects.
Especially, it has been used to study current-induced dynamics in model
systems\cite{brandbyge_theory_1994,brandbyge_electronically_1995,mozyrsky_quantum-limited_2004,mozyrsky_intermittent_2006,bode_scattering_2011,bode_current-induced_2012,thomas_scattering_2012,lopez-monis_limit_2012}.
The nonequilibrium nature of the electronic environment brings in several new
effects that are absent in equilibrium, among which are the modification of 
nuclear potential\cite{hussein_semiclassical_2010,metelmann_adiabaticity_2011,dzhioev_kramers_2011,dzhioev_out--equilibrium_2013}, appearance of non-conservative current-induced forces
and effective magnetic field due to the Berry phase of electrons\cite{lu_blowing_2010,lu_laserlike_2011,lu_current-induced_2015,bode_scattering_2011,bode_current-induced_2012,lopez-monis_limit_2012,lu_current-induced_2012,lu_effects_2015,bustos-marun_adiabatic_2013}.  
These current-induced effects show up already in the linear coupling regime. 
Thus, it is convenient to use Eq.~(\ref{eq:lang2}), and perform mode analysis 
in the frequency domain. Equation~(\ref{eq:lang2}) transforms to
\begin{equation}
	-\omega^2 X(\omega) =-KX(\omega) -\Pi_e^r(\omega) X(\omega) + \chi(\omega),
\end{equation}
where $K$ is the dynamical matrix in the harmonic approximation.
The equation can be solved for $X(\omega)$. But, here instead of solving it, 
it is useful to analyze the structure of $\Pi^r_e$. To this end, we 
split $\Pi^r_e$ into four different contributions
\begin{equation}
	\Pi^r_e = \textrm{Re} \Pi^r_{e,sym} + \textrm{Re} \Pi^r_{e,asym} + i\textrm{Im} \Pi^r_{e,sym} + i\textrm{Im} \Pi^r_{e,asym}.
    \label{eq:pdiv}
\end{equation}
The real part $\textrm{Re}\Pi^r_e$ has a symmetric $\textrm{Re}\Pi^r_{e,sym}$ and an
anti-symmetric $\textrm{Re}\Pi^r_{e,asym}$ part in the vibrational mode index.  The imaginary part $\textrm{Im}\Pi^r_e$ is
similar. The two anti-symmetric parts are nonequilibrium contributions and are
zero in equilibrium. In the following, we analyze these terms one by one.

%----------------------------------
\subsection{Electronic friction}
%----------------------------------
Electronic friction comes out of the adiabatic expansion as the first order
correction yielding $\Gamma_e$ in Eq.~(\ref{eq:lang1}). This dissipative force leads to 
energy transfer between the electrons and the nuclei, thus is beyond the
Born-Oppenheimer approximation. This energy transfer plays an important role in
adsorbate dynamics on metal
surface\cite{headgordon_vibrational_1992,krishna_vibrational_2006,chang_infrared_1990,white_conversion_2005,yang_vibrational_1992,nienhaus_electron-hole_1999,gergen_chemically_2001,trail_energy_2002}.
In our harmonic model with small-$x$ expansion, the third term in
Eq.~(\ref{eq:pdiv}) describes the electronic friction felt by the nuclei. From
it, we can define a frequency-dependent, non-Markovian $\Gamma_e(\omega) =
\textrm{Im}\Pi^r_{e,sym}(\omega)/\omega$. 
Its diagonal elements represent
effective broadening of vibrational modes and are related to the vibrational lifetime.

Physically, the electronic friction originates from excitation of 
EHP in the Fermi sea by nuclear motion. It has been derived using different
theoretical approaches\cite{headgordon_molecular_1995,mozyrsky_intermittent_2006,brandbyge_electronically_1995,brandbyge_theory_1994,lu_blowing_2010,bode_scattering_2011,dou_universality_2017,dou_electronic_2017,dou_born-oppenheimer_2017}.
One notable example is that of Head-Gordon and Tully
\cite{headgordon_molecular_1995,headgordon_vibrational_1992,head-gordon_molecular-orbital_1992}.  
By deriving the nonadiabatic coupling between electronic states and 
applying GLE and mean-field theory, they calculated the electronic friction for the CO/Cu(100)
system. This approach has now become a standard tool
in the study of adsorbate dynamics on metal surface, even under external
driving.  Recently, the tensorial and mode-specific feature of the friction 
matrix in energy transfer has been analyzed which we return to later\cite{askerka_role_2016,maurer_mode_2017}, 
%(see Figs.~\ref{fig:cocu}-\ref{fig:h2ag}). 
% MB: cite figures in sequence
and the similarities and 
differences of different definitions have also been discussed\cite{dou_universality_2017}.
%---------------------------------------------------------------------------------
\begin{figure}[!ht]
\begin{center}
\includegraphics{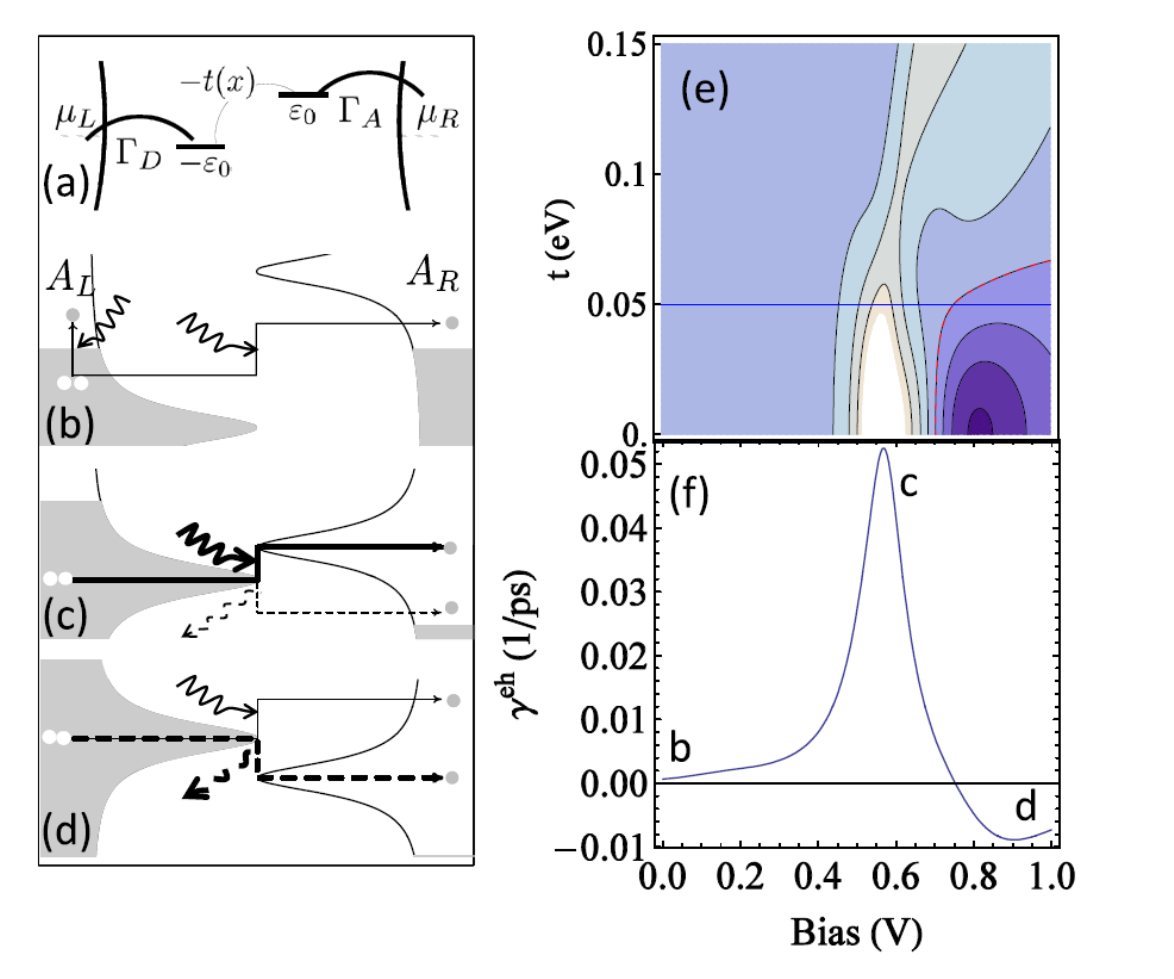}
\end{center}
\caption{(a) Schematics of the a donor-acceptor system coupling to a vibrational mode, whose displacement is denoted by $x$. The donor and acceptor couples to the left and right electrodes, respectively, with broadening parameter $\Gamma_D$ and $\Gamma_A$. (b) Typical EHP excitations at zero bias. Intra-electrode excitation in the left electrode and inter-electrode excitation from the left to the right electrode are shown. Similar excitation in the right electrode and from the right to left are also present, but not shown here. (c) Finite bias, resonant absorption of the vibration. This corresponds to c in (f), where the electronic friction is maximum. (d) Finite bias, resonant emission of the vibration, corresponding to the d in (f). The electronic friction becomes negative. (e) Contour plot of the electronic friction as a function of bias and hopping element $t$ between the donor and acceptor site. The red line separates the negative from the positive friction region. (f) A line cut of (e) at $t=0.05$ eV. Figure adopted from Ref.~\cite{lu_laserlike_2011} with permission.}
\label{fig:PRL11}
\end{figure}

The electronic friction we derived here also applies to the nonequilibrium case, 
i.e., in the presence of electrical current. The electronic density of states (DOS)
determines directly the magnitude of the electronic friction.
If the electronic DOS around the bias window is flat, we can neglect its energy dependence. In this case
 the bias dependence of the electronic friction will be negligible corresponding to 
the wide band limit in quantum transport. In the opposite case,
the bias dependence becomes important and new effect may emerge.
In Ref.~\cite{lu_laserlike_2011}, the authors considered a single vibrational
mode coupling to a donor-acceptor two-level electronic system (Fig.~\ref{fig:PRL11} (a)). 
For this single mode model quantities in Eq.~(\ref{eq:lang1}) are all numbers. 
The bias dependence of the electronic friction $\gamma_e$ can be expressed from 
the rates of vibrational emission ($\mathcal{B}$) and absorption ($\mathcal{A}$)  processes, see Fig.~\ref{fig:PRL11}(b)-(f),
\begin{equation}
	\gamma_e = \mathcal{A} - \mathcal{B}.
	\label{}
\end{equation}
Depending on the relative position of the donor and acceptor level, there could
be resonantly enhanced emission or absorption of the vibrational mode. For
resonant absorption, $\mathcal{A} > \mathcal{B}$, and $\gamma_e$ is large
(Fig.~\ref{fig:PRL11} (c)), while for resonant emission $\gamma_e$ decreases
and even goes negative  (Fig.~\ref{fig:PRL11} (d)). In the former case, the
donor level is lower than the acceptor level by one vibrational quantum
$\Omega$ and the main transport channel is accompanied by vibrational absorption. Thus,
the current can be used to depopulate the vibrational mode, leading to
current-induced cooling of the mode.  On the other hand, in the latter case, the position of the
two levels are reversed.  This population inversion is akin to the case of
the mode populations in a laser. This vibrational amplification by stimulated emission is the physical
reason that leads to the negative electronic friction. This negative friction
is a nonequilibrium effect, contrary to equilibrium, where the relative
magnitude of $\mathcal{A}$ and $\mathcal{B}$ is determined by the detailed
balance relation $\mathcal{B}/\mathcal{A}=e^{-\Omega/k_BT}<1$, resulting in a positive friction. 
The anharmonic effect on heating and cooling of
the molecular junction has been analyzed by Segal and
coauthors\cite{simine_vibrational_2012}.

\begin{figure}[!ht]
\begin{center}
\includegraphics[scale=0.5]{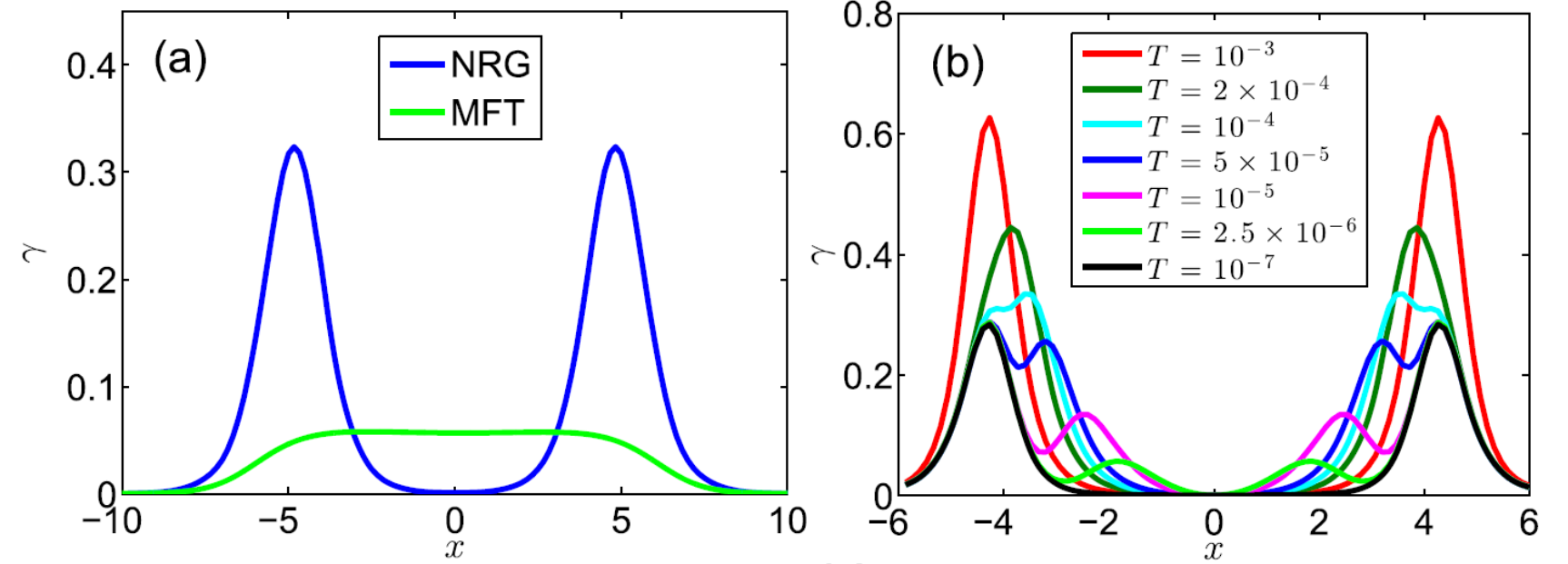}
\end{center}
\caption{Calculation of electronic friction $\gamma$ of an Anderson-Holstein model. (a) Electronic friction $\gamma$ as a function of position $x$ using the numerical renormalization group (NRG) and mean field theory (MFT) calculations. Note that MFT fails to recover two peaks in the friction. (b) Electronic friction according to NRG theory at low temperature; note that the two peaks in friction split into four peaks at low temperature. 
%(c) The Kondo temperature $T_K(x)$ as a function of position and the physical temperature $T(x)$  for which we find a peak in friction at position $x$. Note that these two temperatures are in rough agreement. (d) The height of the Kondo peak $\gamma_k$ as a function of temperature.  Note that these peaks decrease exponentially and vanish at zero temperature.  
Figure adopted from Ref. \cite{dou_born-oppenheimer_2017} with permission. }
\label{fig:dou17}
\end{figure}

In our theory, we have ignored the effect of electron-electron 
interaction on the electronic friction beyond the adiabatic mean-field screening of the coupling.
Combined with numerical renormalization group calculation, 
Dou \emph{et al}.\cite{dou_born-oppenheimer_2017} studied the modification of electronic friction due to strong 
electron-electron correlation through the Anderson-Holstein model. 
They found a qualitative difference of the electronic friction calculated 
from the numerical renormalization group (NRG) and the dynamical mean field theory (MFT) (Fig.~\ref{fig:dou17}).
This highlights the importance of electron correlation on the electronic friction.

%------------------------------------------------------------------------------
\subsection{Joule heating from the nonequilibrium fluctuations}
%------------------------------------------------------------------------------
\label{subsec:joule}
The interaction of the flowing current with the nuclei leads to heat transfer
from the electronic to the nuclear DoF, normally termed Joule heating. In the
SGLE, this is reflected in the correlation function of the fluctuating force.
In equilibrium, we have the fluctuation-dissipation relation
\begin{equation}
	\langle \chi_e(\omega) \chi^*_e(\omega)\rangle_{\rm eq}  = -\textrm{Im} \Pi_e^r(\omega) \coth(\omega/2k_BT).
    \label{eq:fd}
\end{equation}
This includes both thermal and quantum fluctuations. In nonequilibrium, the
correlation function of $\chi_e$ gains an extra term due to the voltage bias. In the
wide band limit, it has a bias and energy dependence as
\begin{equation}
	\delta \langle \chi_e(\omega) \chi^*_e(\omega)\rangle  \propto (eV-\omega)\Theta(eV-\omega).
	\label{}
\end{equation}
The extra noise is linear in $\omega$. The Heaviside step function $\Theta(eV-\omega)$ means that,
due to energy conservation,
the extra noise have an upper limit determined by the applied bias.
Fitting the nonequilibrium noise correlation function to a form similar to Eq.~(\ref{eq:fd}),
we can define an effective electronic temperature of a given vibrational mode $i$
in the presence  of current flow,
\begin{equation}
	\langle \chi_{e,i}(\omega) \chi^*_{e,i}(\omega)\rangle_{\rm neq}  = -\textrm{Im} \Pi_{e,ii}^r(\omega) \coth(\omega/2k_BT_{{\rm eff},i}).
    \label{eq:fd11}
\end{equation}
It should be noted that, the above equation contains the electron-nuclear coupling terms and different modes will generally experience different effective electronic temperatures.
From this analysis, Joule heating can be understood from another point of view. 
The applied bias changes the effective temperature of the electronic system and
the temperature difference leads to heat flow between the electrons and the nuclei.

It can be shown analytically that, for harmonic oscillators, using the above noise correlations, the
prediction of Joule heating from the SGLE is equivalently to that from the NEGF method
under the same approximations\cite{lu_current-induced_2012}. That is, 
it can produce fully quantum mechanical results for harmonic oscillators. 

%-----------------------------------------------------------------------------------------------------------------
\subsection{The non-conservative and  effective Lorentz force}
%------------------------------------------------------------------------------------------------------------------

\begin{figure}[!ht]
\begin{center}
\includegraphics{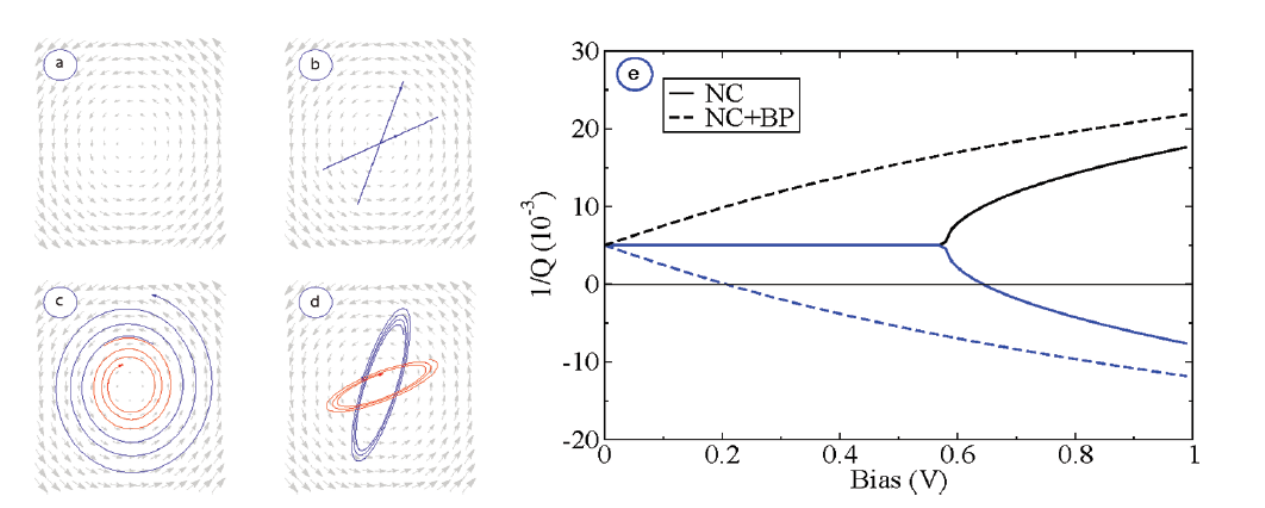}
\end{center}
\caption{(a) Schematics of the force field generated by the non-conservative current-induced forces.
(b) Trajectories of two linear harmonic oscillators without non-conservative forces. 
(c) Trajectories in the presence of non-conservative forces. The amplitude of the red mode damps with time, 
while that of the blue mode grows with time, indicating an instability. 
(d) When both the non-conservative and the Lorentz force are present, the trajectories changes. 
(e) Inverse $Q$ factor of the two modes as a function of bias in the presence of non-conservative (NC) or Lorentz (BP) force. Negative $1/Q$ means instability indicated in (c) and (d). Figure adopted from Ref.~\cite{lu_blowing_2010} with permission.}
\label{fig:NL10}
\end{figure}

When a current flows through a conductor, it induces forces on the nuclear DoF.
Whether this current-induced force is conservative or not is a question that
brought some confusion\cite{landauer_driving_1974,sham_microscopic_1975,sorbello_local_1977,di_ventra_are_2004}. Todorov and co-authors gave a concise answer to this
question\cite{dundas_current-driven_2009,todorov_nonconservative_2010,todorov_nonconservative_2011}.  Moreover, they showed that
the non-conservative force can be used to drive an atomic motor using Enhrenfest MD to
perform the numerical calculations\cite{dundas_current-driven_2009,mceniry_modelling_2010}. \revision{Later on, the SGLE was
used to study the same problem\cite{lu_blowing_2010}, and
the effect has been extended to mesoscopic systems\cite{bode_scattering_2011,bustos-marun_adiabatic_2013,fernandez-alcazar_decoherence_2015,fernandez-alcazar_dynamics_2017,calvo_real-time_2017}.} It
has the advantage  of considering the deterministic current-induced forces and
stochastic Joule heating on an equal
footing\cite{lu_blowing_2010,lu_current-induced_2012,bode_scattering_2011,bode_current-induced_2012,thomas_scattering_2012}.
It was here predicted that, in additional to the non-conservative force, there is an extra
effective Lorentz force originating from the Berry phase of the electrons. 
We discussed how the Joule heating can be considered within the SGLE in
Subsec.~\ref{subsec:joule} in terms of stochastic forces, while here we focus on the deterministic current-induced
forces.  

Firstly, the force contributed from $F_{nc}\equiv -\textrm{Re}\Pi^r_{e,asym}X$ is
non-conservative.  This can be seen from the anti-symmetric properties of
$\textrm{Re}\Pi^r_{e,asym}$, which leads to  $\nabla \times F_{nc}\neq 0$. This
means, the nuclei move within a non-conservative force field
(Fig.~\ref{fig:NL10} (a)). If they move along a certain loop, $F_{nc}$ can pump
or extract energy from the nuclei depending on the direction of the motion
(Fig.~\ref{fig:NL10} (c, d)). This energy transfer through deterministic work
is fundamentally different from Joule heating.  Since the off-diagonal (anti-symmetric) part of
$\textrm{Re}\Pi^r_{e,asym}$ is important, the system should have at least two DoF.

Secondly, the force contributed from $F_{bp}\equiv -\textrm{Im}\Pi^r_{e,asym}\dot{X}$
is different from friction. Its effect on the nuclear dynamics is similar to that of an
magnetic field, due to the anti-symmetric property of $\textrm{Im}\Pi^r_{e,asym}$.  It
originates from the Berry phase of the electrons, which back acts on the
nuclei. Actually, this force is more easily understood from the adiabatic
expansion. It, together with the friction, comes as the first order 
correction in the expansion($-\Gamma_e \dot{X}$ term in Eq.~(\ref{eq:lang1})).
We should mention that the Berry phase comes from the time-reversal symmetry
breaking in the electronic environment, and becomes zero in equilibrium. This
phase is not quantized, and $F_{bp}$ changes continuously with the applied voltage.
Berry and Robbins have studied this kind of `geometric magnetism' and showed that it is
zero when the system has a discrete spectrum\cite{berry_chaotic_1993}.  Here,
the coupling of the system to the electronic environment leads to broadening of
the spectrum, and renders the Berry phase non-zero. Like the Lorentz force, 
$F_{bp}$ does no work on the nuclei, but it changes the orbit of the eigen mode motion. 
It may change the orbit from linear to elliptical (Fig.~\ref{fig:NL10}), and help the
non-conservative force to do work on the nuclei.

Both $F_{nc}$ and $F_{bp}$ come into play only when there are at least two DoF
in the system. Thus, the main results can be illustrated by a two-mode
model\cite{lu_blowing_2010,lu_current-induced_2012}.
Neglecting the fluctuating forces first, we can write their equations of motion
in the eigen mode basis
\begin{equation}
\left(\begin{array}{c}
\ddot{x}\\ \ddot{y} 
\end{array}\right)
=-\left(\begin{array}{cc}
\omega_1^2 & a\\ -a & \omega_2^2
\end{array}\right) 
\left(\begin{array}{c}x\\ y\end{array}\right) -
\left(\begin{array}{cc}
\gamma & b\\ -b & \gamma
\end{array}\right) 
\left(
\begin{array}{c}
\dot{x}\\ \dot{y}
\end{array}
\right).
\end{equation}
Here, $\omega_1$ and $\omega_2$ are the angular frequency of the two otherwise independent vibrational modes.
Their simultaneous coupling to the electrical current leads to indirect vibrational coupling, 
parametrized by $a$ and $b$ in the above equation. They represent the non-conservative and
the effective Lorentz force, respectively. 
We can analyze the eigen spectrum of the system including $a$ and $b$. 
We find that, above certain bias, the two forces, especially $F_{nc}$, 
may drive the system into a run-away instability,
characterized by a negative damping corresponding to a negative $1/Q$ factor, as shown in Fig.~\ref{fig:NL10} (e). In this situation the energy of the unstable harmonic mode will keep increasing in time once it is excited.  
It is also shown that, the change of the orbit by $F_{bp}$ helps $F_{nc}$ to perform
work. This is f.ex. seen in how the threshold bias of the instability decreases when including $F_{bp}$.

%-----------------------------------------------------------------------------------------
\subsection{Renormalization of the vibrational potential and bistability}
%-----------------------------------------------------------------------------------------
\begin{figure}[!ht]
\begin{center}
	\includegraphics[scale=0.6]{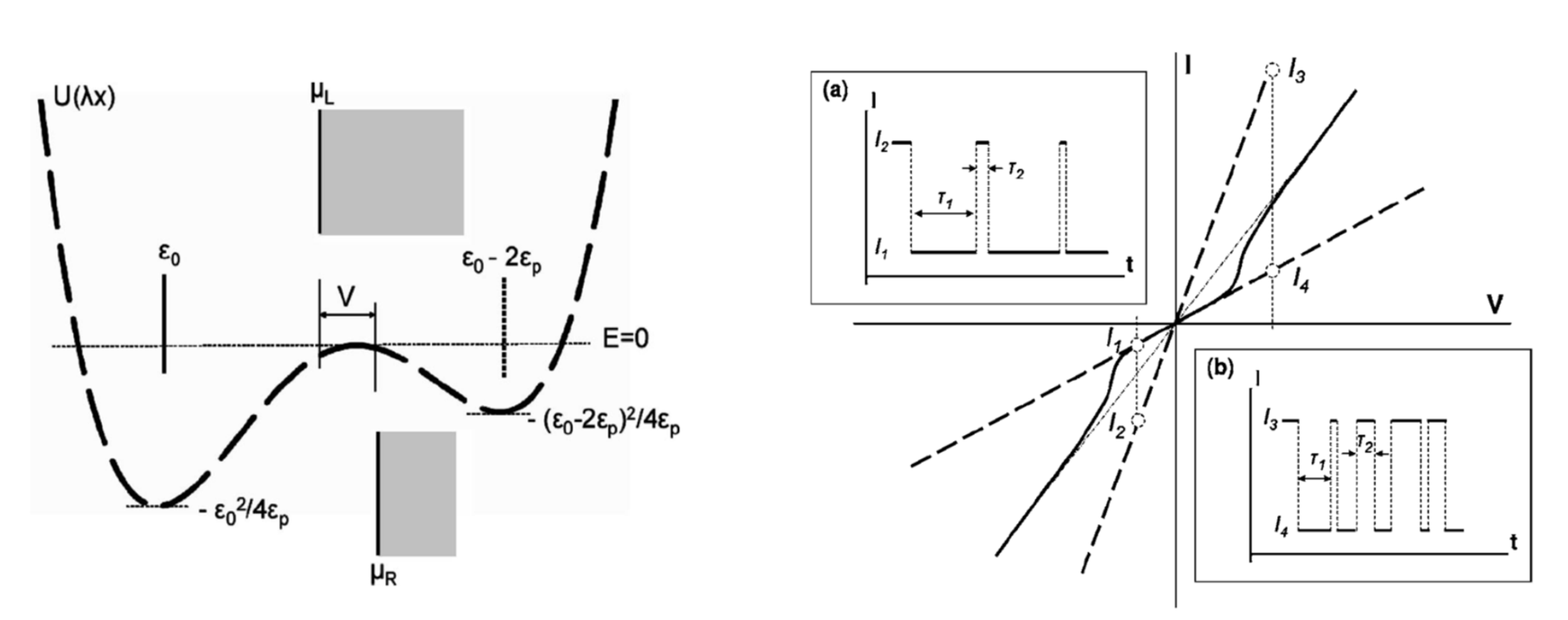}
\end{center}
% \caption{Strong electron-vibration interaction leads to a bistable potential. }
\caption{Left: The potential profile of a harmonic oscillator coupling to one single electronic level at finite bias. The single well harmonic potential is modified and acquires a bistable profile at finite bias. 
Right: The $I$-$V$ curve of the single level model showing bistability. (a) The temporal dependence of the electrical current at small bias, where the oscillator spends most of the time in a deeper well and occasionally jumps to a shallower well. (b) The same as (a), but at higher voltages. Now the probability to occupy the shallower well, and the switching rate increase due to higher effective temperature. Figure adopted from Ref. \cite{mozyrsky_intermittent_2006} with permission. 
}
\label{fig:mozy06}
\end{figure}
\begin{figure}[!ht]
\begin{center}
\includegraphics[scale=0.6]{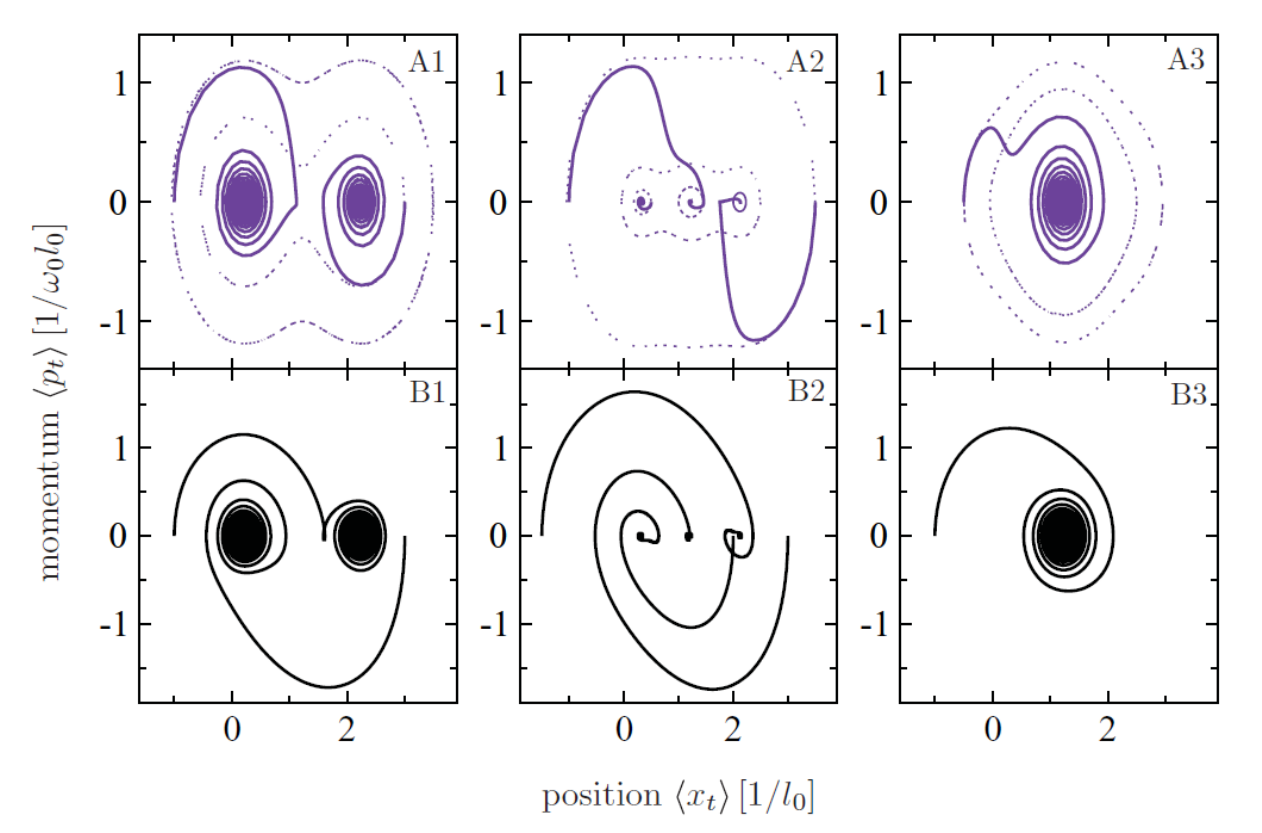}
\end{center}
\caption{Bias dependent phase space trajectories for a single electronic state coupling to a vibrational mode in the Holstein form. Dashed lines in row A are results from
the Born-Oppenheimer dynamics without friction, solid lines include friction. Row B shows results including all the non-adiabatic effects in the SGLE.
The voltage bias modifies the number of fixed points and the trajectories drastically.
Figure adopted from Ref. \cite{metelmann_adiabaticity_2011} with permission. 
}
\label{fig:brandes11}
\end{figure}

If the nuclei couple strongly to the electrons, the potential felt by the
nuclei may change a lot when there is an applied voltage bias. It may even
generate a bistable state and lead to conductance switching behavior. 
It has been studied within the minimal Anderson-Holstein
model\cite{mozyrsky_intermittent_2006,metelmann_adiabaticity_2011},
where one vibrational mode couples to the electronic system through an onsite interaction.
The left part of Fig.~\ref{fig:mozy06} shows the voltage induced bistable states.
It comes from a harmonic potential at zero bias. The right
part of Fig.~\ref{fig:mozy06} shows its signature in the electrical current. 
In Fig.~\ref{fig:brandes11}, different kinds of behaviors in
phase space are also observed at finite biases. 

%%%%%%%%%%%%%%%%%%%%%%%%%%%%%%%%%%%%%%
\section{Applications}
\label{sec:app}
%%%%%%%%%%%%%%%%%%%%%%%%%%%%%%%%%%%%%%
\subsection{Numerical implementation}
Before turning to the applications of the SGLE, we discuss briefly two technical
issues faced in any numerical implementation of the SGLE approach. 
Firstly, the friction part of the SGLE in general has a memory kernel due to
the difficulty in separating completely the time scales of the system and the 
bath. Although this does not pose a conceptual problem, in practical, realistic
calculations, it is numerically more involved compared to a time-local friction.  
To this end the non-local, non-Markovian equation can be transformed into a local,
Markovian one by introducing auxiliary DoF into the equation. This trick
has been successfully used by different 
authors \cite{ceriotti_colored-noise_2010,stella_generalized_2014}.
Secondly, the colored noise spectrum requires certain attention in the generation
of the random force.  Two approaches have been applied in the literature.
This first one makes use of the fact that the noise is $\delta$-correlated in the frequency
domain. Thus, generating the noise in the frequency domain and making fast
Fourier transform to time domain gives a time-correlated noise that can be used to do
the MD simulation\cite{lu_numerical_2005,wang_quantum_2007}. 
The problem of this method is that, one has
to generate the noise and store it before performing the simulation. For large
scale simulations involving many DoF this could result in a storage problem. More importantly, for certain
applications, i.e. molecular diffusion on, or scattering off, a metal surface, 
the coupling of the system to the baths may change during the simulation. 
In this case one needs to adjust the noise on the fly.  To overcome this difficulty, 
different approaches
have been adopted to generate the noise directly in time domain \cite{barrat_portable_2011,stella_generalized_2014,schmidt_simulation_2015}.
%iterative reconstruction of the memory kernel \cite{jung_iterative_2017}

%++++++++++++++++++++++++++++++++++++++++++++++++++
%\subsection{Electronic friction}
\subsection{Adsorbate dynamics on metal surfaces}
%++++++++++++++++++++++++++++++++++++++++++++++++++

The coupling between molecular motion and EHP excitation 
is important in order to understand adsorbate dynamics at surfaces,
F. ex. it was found that the vibrational lifetime of CO on a metal surface is in the
order of ps, while on a NaCl surface it is many orders of magnitude longer
(ms)\cite{chang_infrared_1990,morin_vibrational_1992}, \revision{see Ref.~\cite{alducin_non-adiabatic_2017} for a recent review}.  Detailed theoretical
and experimental study shows that the coupling between molecular vibrations and
the EHP excitation is key to understand this difference.
Molecular-beam surface scattering experiments provide direct evidence of
the coupling of molecular motion to the EHP
excitation\cite{rettner_observation_1985,rettner_direct_1987,white_conversion_2005,yang_vibrational_1992,yang_controlling_1993}.
The detection of chemicurrent during the adsorption of molecules on thin
metal surface of tunnel junctions is another signature of coupling\cite{nienhaus_electron-hole_1999,gergen_chemically_2001,trail_energy_2002}.
The electronic friction language was quite successful in understanding these
experiments.  Head-Gordon and Tully \cite{headgordon_molecular_1995} in 1995
developed a method to perform classical MD including electronic
friction from metal electrons. By deriving the nonadiabatic coupling between
electronic states and applying the GLE, they obtained electronic friction for 
the CO/Cu(100) system. Since then, generalized Langevin equations has become 
an important tool in the study of adsorbate
dynamics on metal surface, even under external driving. 

Although the general picture discussed above is widely accepted, obtaining quantitative measures of
 the role of EHP excitation in the adsorption and reaction
dynamics is still challenging.
Here, we only summarize recent theoretical and experimental advances in typical
systems\cite{vazhappilly_femtosecond-laser_2009,fuchsel_concept_2012,blanco-rey_electronic_2014,rittmeyer_electronic_2015,askerka_role_2016,maurer_mode_2017,galperin_nuclear_2015,novko_ab_2015,grotemeyer_electronic_2014,dou_frictional_2015,dou_universality_2017,dou_electronic_2017,dou_born-oppenheimer_2017,fuchsel_enigmatic_2016,kruger_no_2015,scholz_femtosecond-laser_2016,luo_electron-hole_2016,kroes_vibrational_2017}.

%---------------------------------------------------------------------------------
%Example 1.
%H atom scattering on Au(111) surface
\begin{figure}[!ht]
	\begin{center}
	\includegraphics[scale=0.6]{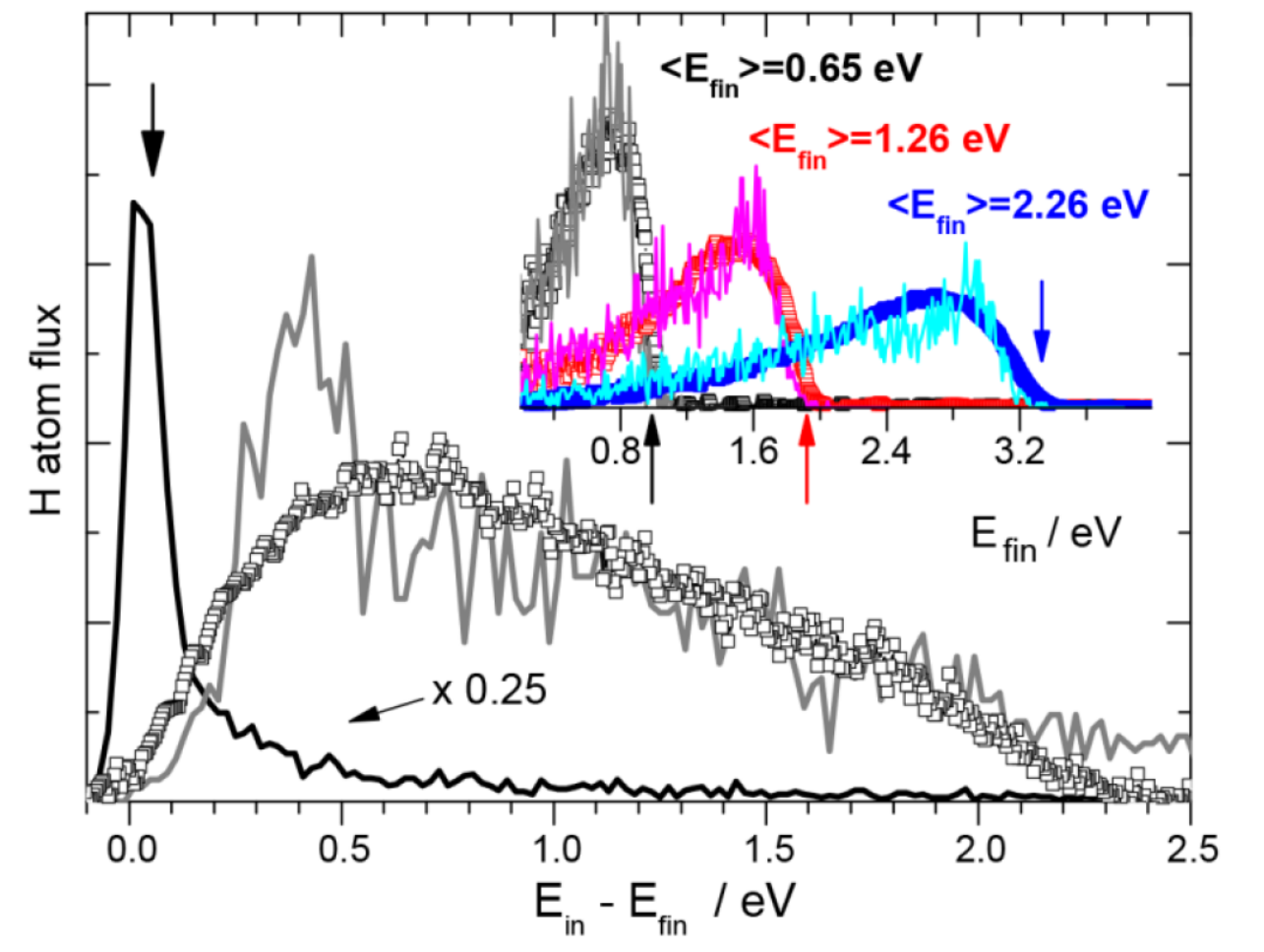}
	\end{center}
% \caption{H atom scattering onto Au(111) surface. Science, 350, 1346 (2015). }
\caption{Comparison of the experimentally obtained kinetic energy loss 
spectrum to theoretical simulations. Theoretical energy loss found when 
neglecting (solid black line), and including (solid gray line) electronic 
excitation. Experimental energy loss for $E_{in}$ = 2.76 eV is shown as open 
squares. The vertical arrow marks the expected energy loss for a binary 
collision between an H and an Au atom. The inset shows the incidence energy 
dependence, $E_{in}$, of the experimentally derived translational inelasticity (open squares) and comparison to theory (solid lines): $E_{in}$  = 3.33 eV (blue), 1.92 eV 
(red), and 0.99 eV (black). Colored arrows mark the three incidence 
energies. Also shown are the average final translational energies, $\langle E_{fin} \rangle$. The 
scattering angles are 
$\theta_i= 45°$ , 
$\theta_s= 45°$ , and 
$\phi_i= 0°$  with respect to the [101] 
direction. In all cases, the scattered H atoms remain unthermalized with the 
solid, emerging with a substantial fraction of their incidence translational 
energy. Figure adopted from Ref. \cite{bunermann_electron-hole_2015} with permission. }
% Note: incident energy $E_{in}$, final energy $E_{fin}$
\label{fig:hau}
\end{figure}
%H atom scattering on Au(111) surface
B\"unermann \emph{et al}. studied inelastic H-atom scattering from Au(111)
surface\cite{bunermann_electron-hole_2015}.  A beam of nearly mono-energetic H
atoms was prepared by laser photolysis and injected on to Au(111) surface with
and without adsorbed Xe layers. The experimental setup allowed study of
inelastic H-atom scattering with different incidence and scattering angles. The
experimental results show drastic difference between Xe and Au surface
scattering. For Xe surface, due to the requirement of energy and momentum
conservation, the translational energy loss of the H-atom was small. For
Au(111) surface, the translational energy of H-atom can be directly converted
to EHP excitations. Thus, the energy loss was much larger than in
the Xe surface case. The experimental results were compared to theoretical simulations
with and without including the electronic friction (Fig.~\ref{fig:hau}), and it was seen how the electronic friction was needed to obtain a reasonable agreement
with the experimental results. This work shows clear evidence of translational
energy dissipation due to electronic friction.
%---------------------------------------------------------------------------------

%---------------------------------------------------------------------------------
%Example 2: Mode specific electronic friction in dissociative chemisorption on metal
%surfaces: H2 on Ag(111)
\begin{figure}[!ht]
	\begin{center}
	\includegraphics[scale=0.6]{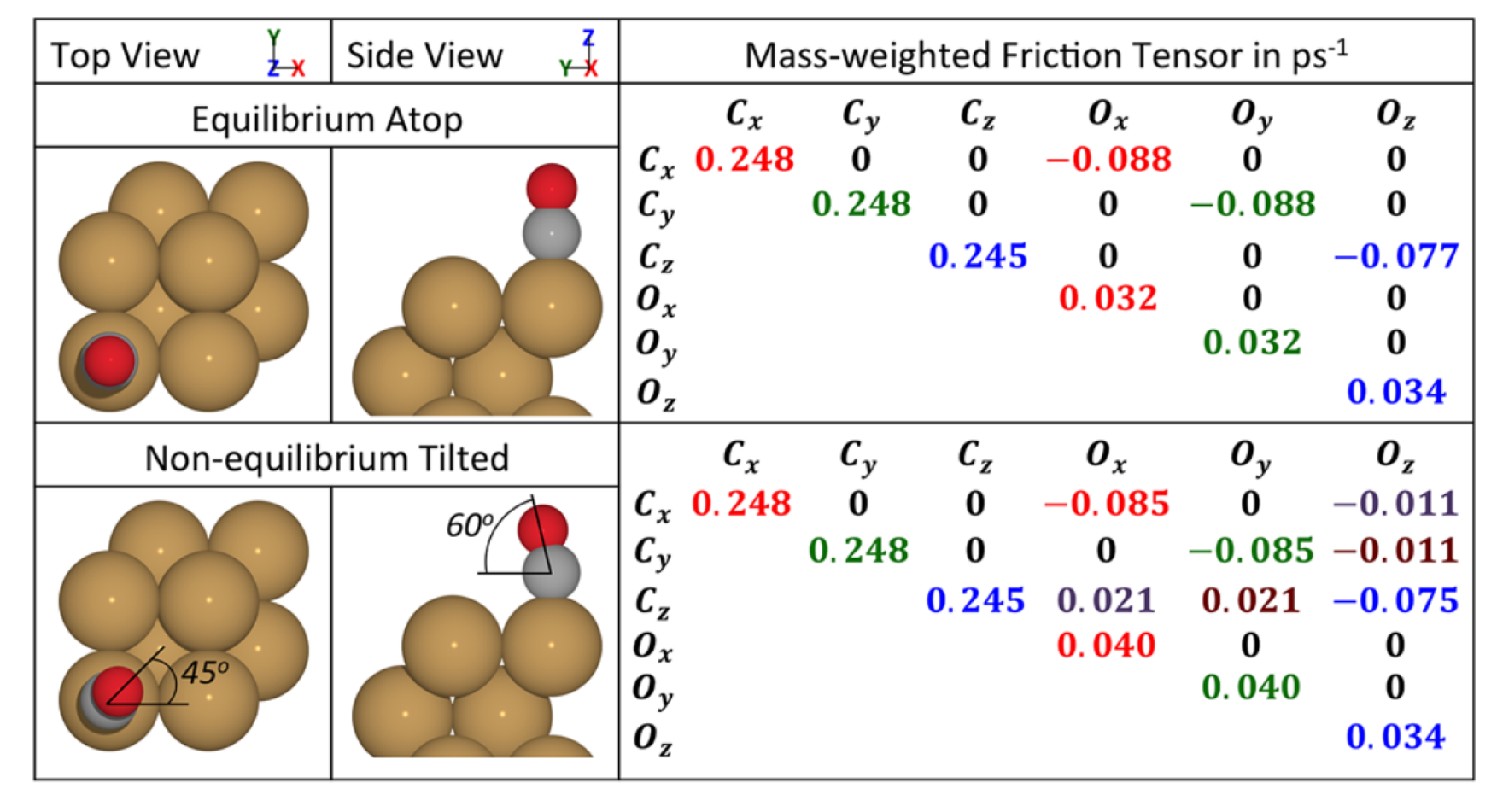}
	\end{center}
\caption{CO on top of Cu(100) surface. (left) CO adsorbed at an atop site of a Cu(100) surface in the equilibrium position and a nonequilibrium tilted geometry as viewed from $xy$ (top view) and $yz$ (side view) planes. (right) Mass-weighted friction tensor (in ps$^{-1}$) for the two geometries. The coloring scheme is consistent for $x$ (red), $y$ (green), $z$ (blue), $xz$ (purple), and $yz$ (brown) components of the friction tensor. Components smaller than $0.002$ ps$^{-1}$ are set to zero. Figure adopted from Ref. \cite{askerka_role_2016} with permission. }
\label{fig:cocu}
\end{figure}
\begin{figure}[!ht]
	\begin{center}
	\includegraphics[scale=0.4]{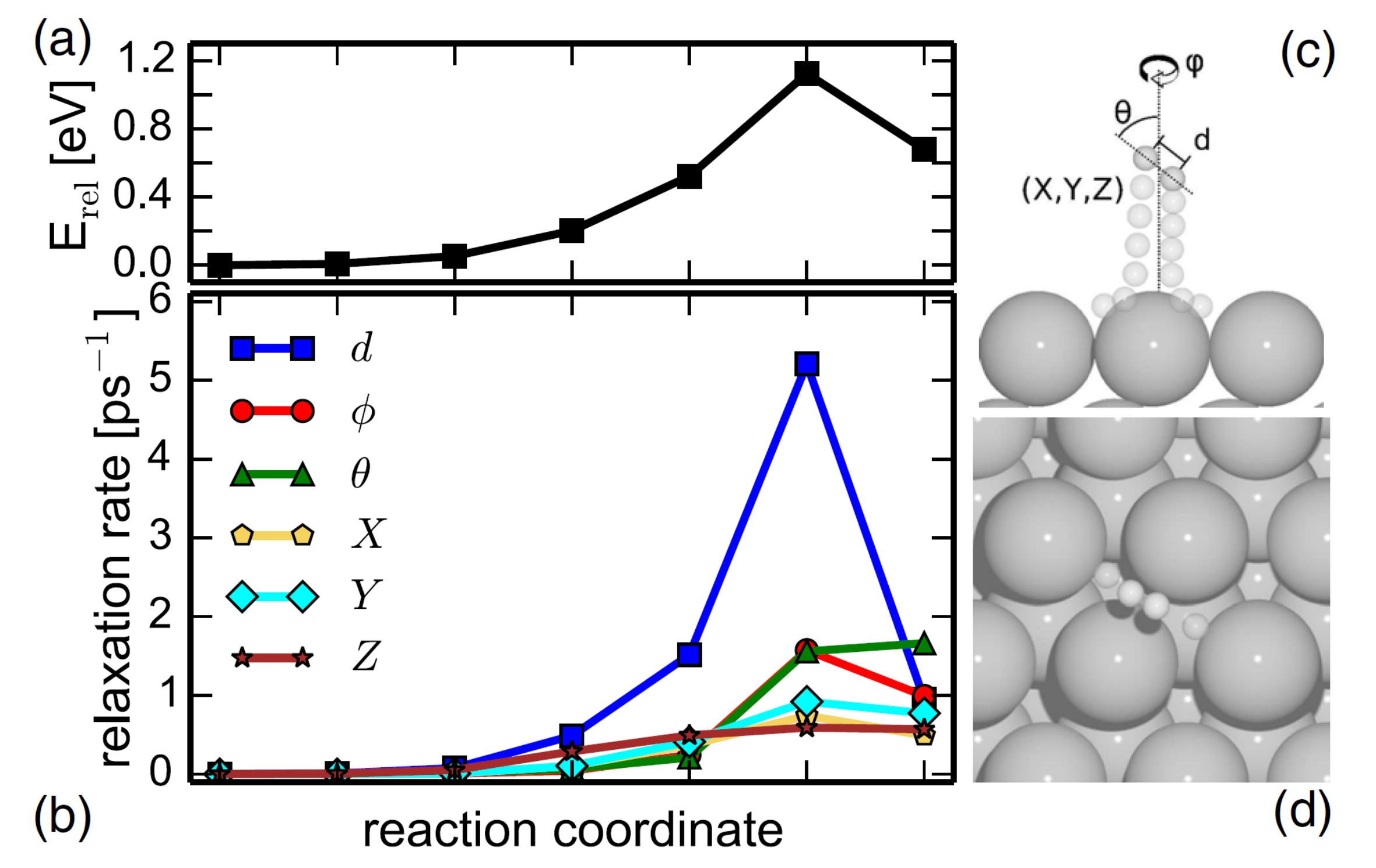}
	\end{center}
% \caption{H2 on Ag(111), PRL, 118, 256001 (2017), Fig. 1.}
\caption{H$_2$ on top of Ag(111).
(a) Relative energy along a minimum energy path (MEP) of H$_2$ dissociation on Ag(111). The path is shown in (c) and (d). 
(b) Relaxation rates in ps$^{-1}$ obtained from the diagonal elements of electronic friction tensor in internal coordinates: bond stretch $d$, azimuthal angle $\phi$, polar angle $\theta$, and the three Cartesian center of mass coordinates of the hydrogen molecule $(X,Y,Z)$. Off-diagonal elements (not shown) can modify relaxation rates by 30\% or more. 
(c) Side view of the dissociation along the minimum energy path. 
(d) Top view of start and end point of MEP. Figure adopted from Ref. \cite{maurer_mode_2017} with permission.}
\label{fig:h2ag}
\end{figure}

The tensorial feature of $\Gamma_e$ matrix in Eq.~(\ref{eq:lang1}) was
emphasized in recent works\cite{askerka_role_2016,maurer_mode_2017}. The
off-diagonal elements of $\Gamma_e$ in the mode space couple different modes
together and influence the energy distribution within different vibrational
modes. The electronic friction is normally anisotropic, i.e., depending on the
direction in real space and having different magnitude for different
vibrational modes.  Figure~\ref{fig:cocu} shows an example of CO adsorbed at
an atop site of Cu(100) surface. The friction matrix elements in real space are
shown. Based on this understanding, Maurer \emph{et al}. studied the scattering
and dissociative chemisorption of H$_2$ on the Ag(111) surface (Fig.~\ref{fig:h2ag}). 
Although the electronic friction only accounts about 5\% of the energy loss, the anisotropy
of the friction induces dynamical steering that changes the reaction outcomes.

%---------------{}------------------------------------------------------------------

\begin{figure}[!ht]
	\begin{center}
	\includegraphics[scale=0.2]{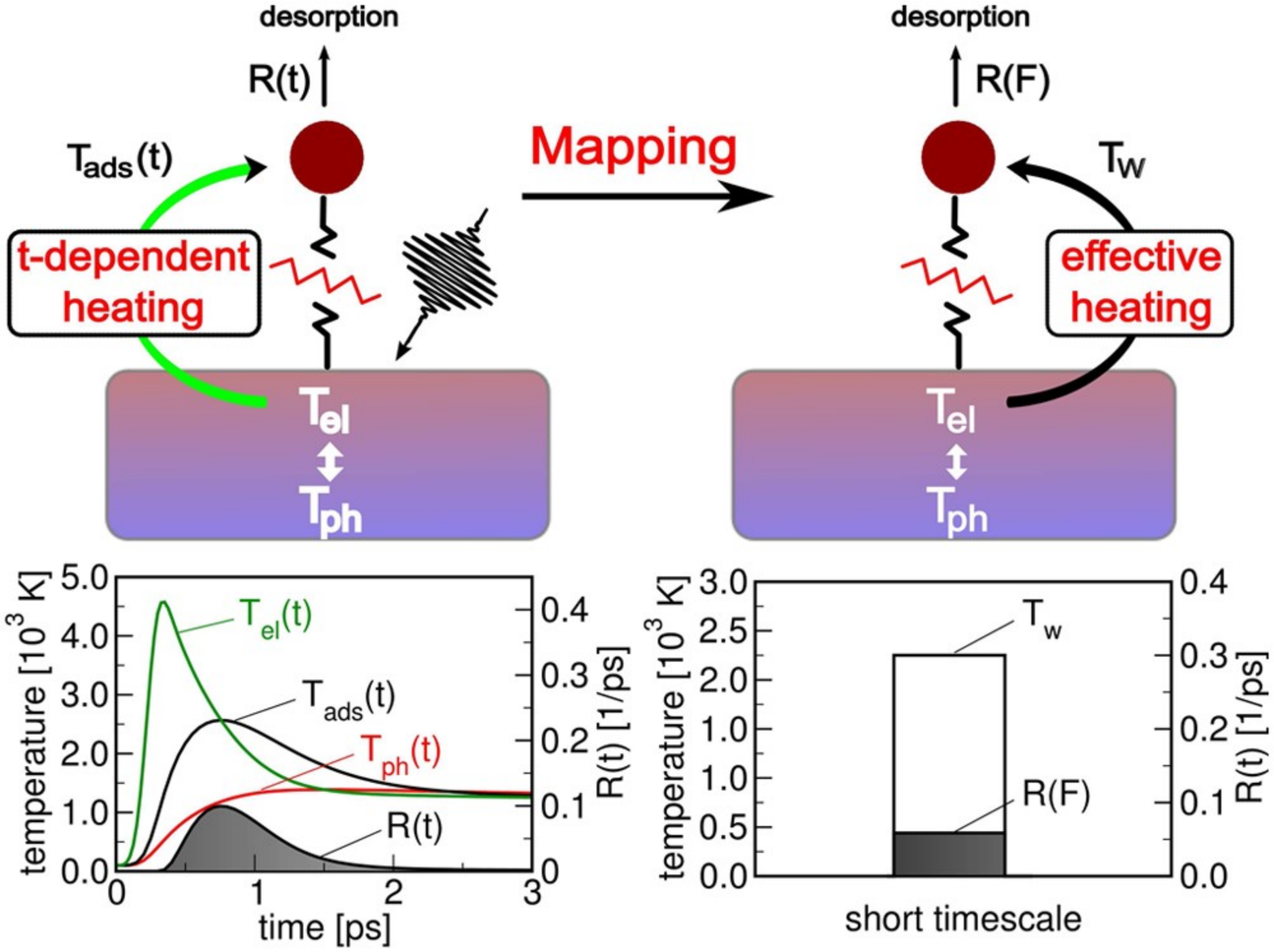}
	\end{center}
% \caption{Desorption process of $D_2$ from $Ru$ after excitation with a 130 fs Gaussian Laser pulse.}
\caption{
Desorption of D$_2$ from Ru after excitation with a 130 fs Gaussian laser pulse at 800 nm, $F =140$ J/m$^2$. 
Left panel: time-dependent behavior of the electron, phonon, and adsorbate temperatures ($T_{el},T_{ph},T_{ads}$). 
% from a 2TM or 3TM, and the desorption rate from the 3TM according to Eq. ?????. 
Right panel: the hot electron-mediated dynamics is mapped to an effective heating mechanism similar to Eq.~\ref{eq:fd11}. $T_w$ represents an effective temperature that captures the physics of the adsorbate excitation, and the fluence $F$ is related to the temperature $T_w$. $T_w$ determines the microscopic properties of the desorbed molecules, and it's of great benefit to numerical simulation. Figure adopted from Ref. \cite{fuechsel_concept_2012} with permission.}
\label{fig:fs_pulse}
\end{figure}

\begin{figure}[!ht]
	\begin{center}
	\includegraphics[scale=0.2]{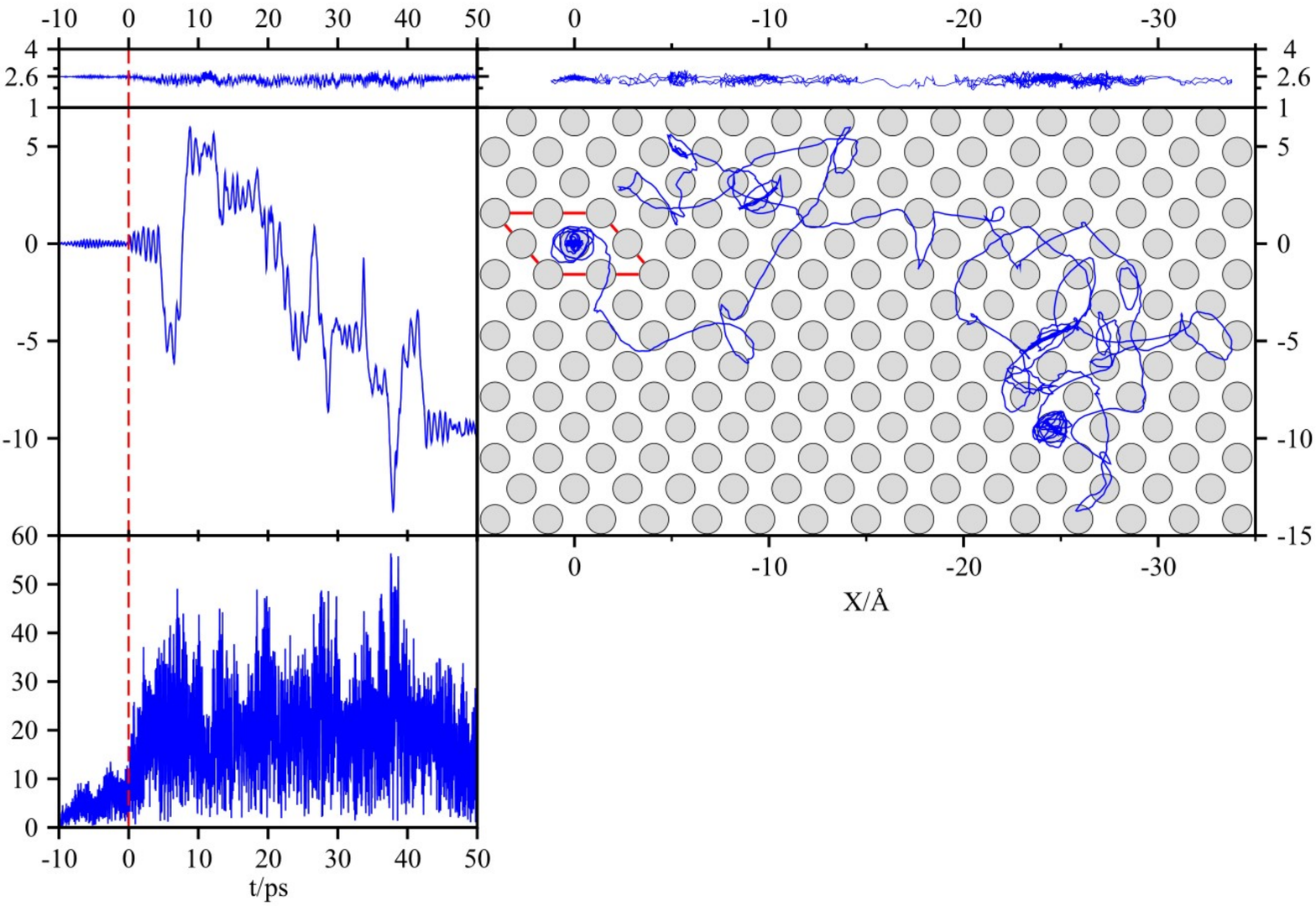}
	\end{center}
% \caption{Time evolution of a typical trajectory of a single CO molecule within the MDEF model.}
\caption{
Time evolution of a typical trajectory of a single CO molecule on Ru(0001) surface from MD with electronic friction, starting from its global minimum, i.e., on
top of a Ru atom (X = Y = 0, $\theta = \phi = 0$), when driven by hot electrons generated from a 140 J/m$^{2}$ laser pulse (at t = 0), $\lambda = 400 nm$, FWHM
= 50 fs, and base temperature 300 K. The top and middle left panels show the center of mass coordinates Z and Y over time, the top right and middle panels
the entire trajectory in the (X,Z) and (X,Y) planes, respectively. The lower left panel gives the tilt angle $\theta$ as a function of time. The top Ru
is visualized with gray circles. 
%Note, only a single
%moving CO is shown and the PES is translationally invariant according to $1 \times 1$ periodic boundaries. 
Figure adopted from Ref. \cite{scholz_femtosecond-laser_2016} with permission.
}
\label{fig:CO-MDEF}
\end{figure}

%---------------------------------------------------------------------------------
%Example 3.
Irradiation of a metal with an ultrafast femtosecond laser pulse induces electronic
excitations and adsorbate desorption from the metal surface. This probes how hot excited electrons
transfer energy to the adsorbate vibrations. The Langevin equation has been used to study these 
processes\cite{fuechsel_concept_2012,scholz_femtosecond-laser_2016}. 
Figures~\ref{fig:fs_pulse} and \ref{fig:CO-MDEF} show two examples of 
simulations after laser excitation.

%---------------------------------------------------------------------------------

%++++++++++++++++++++++++++++++++++++++++++++++++++++++++++++++++++++++
\subsection{Joule heating and current-induced forces in molecular conductors}
%+++++++++++++++++++++++++++++++++++++++++++++++++++++++++++++++++++++++
\begin{figure}[!ht]
	\begin{center}
	\includegraphics[scale=0.6]{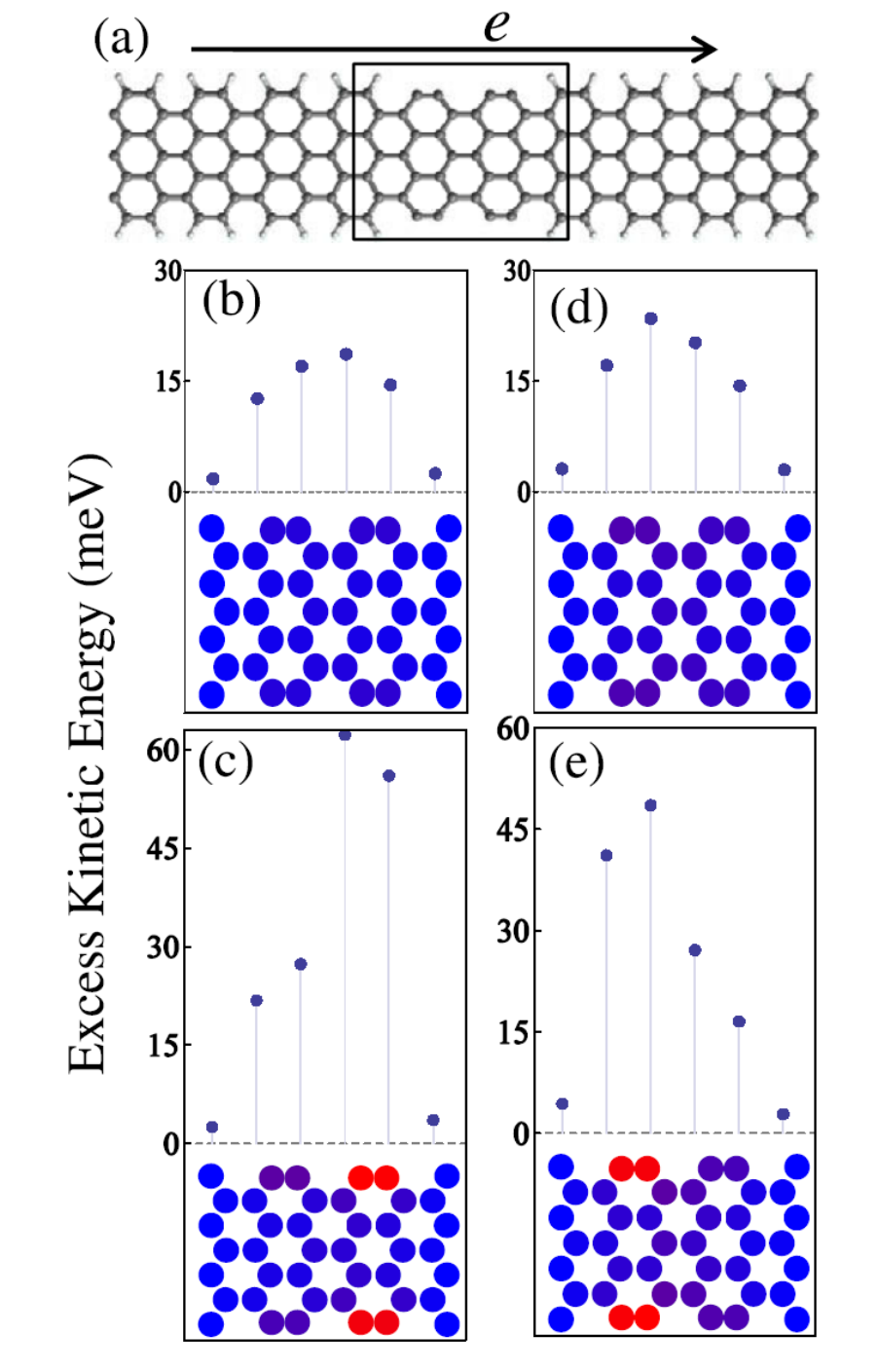}
	\end{center}
% \caption{NC force}
\caption{
(a) Structure of a partially passiviated armchair graphene nanoribbon. The atoms in the solid square couple to a nonequilibrium electronic reservoir.
(b)-(c) The excess kinetic energy of each atom without and with the 
current-induced non-conservative and Lorentz force. The voltage bias is $V=0.4$ V,
the temperature is $T=300$ K, and the Fermi energy $E_F=1.4$ eV. (d)-(e) Same
as (b)-(c), but at $E_F=-1.0$ eV. Figure adopted from Ref. \cite{lu_current-induced_2015} with permission.
}
\label{fig:PRL15}
\end{figure}

One advantage of the SGLE is that it allows consideration of Joule heating and
current-induced forces on an equal footing. The authors in
Ref.~\cite{lu_current-induced_2015} studied the interplay of these two channels of
energy transfer from the nonequilibrium electronic to the nuclear DoF. The
coupling of the system to phonons and electrons in the two electrodes was
considered simultaneously.  It was found that, the effect of the current-induced
force depends on the direction of the current flow. In particular, for a symmetric system with current flow, a asymmetrically placed hot-spot in the nuclear energy distribution (Fig.~\ref{fig:PRL15}) was observed. This effect mainly 
comes from the non-conservative part of the force and depends on the wave 
length of the electron scattering states. These results show the 
important role played by  the non-conservative current-induced force and its relation to the momentum transfer from charge carriers to the nuclear DoF\cite{todorov_nonconservative_2011}. Asymmetric heating has been observed in nanojunctions. These effects are central for the directed motion of atoms in 
the electromigration of nanojunctions\cite{tsutsui_unsymmetrical_2012,schirm_current-driven_2013}.

%++++++++++++++++++++++++++++++++
%\subsection{Current-induced molecular dynamics}
%++++++++++++++++++++++++++++++++
%relation to Enhrenfest dynamics
Our analysis so far is limited to the harmonic approximation for the
vibrations. In the SGLE, the anharmonic interactions is taken into account
classically.  Direct MD simulation using either
Eqs.~(\ref{eq:lang1}-\ref{eq:lang4}) takes the anharmonic
vibrational interactions into account in the same way as the classical MD. But the inclusion
of zero point motion and correct quantum distributions in the SGLE extends its
range of validity to the study of heat transport at low temperature and dynamics in the
presence of electrical current.

\begin{figure}[!ht]
	\begin{center}
	\includegraphics[scale=0.4]{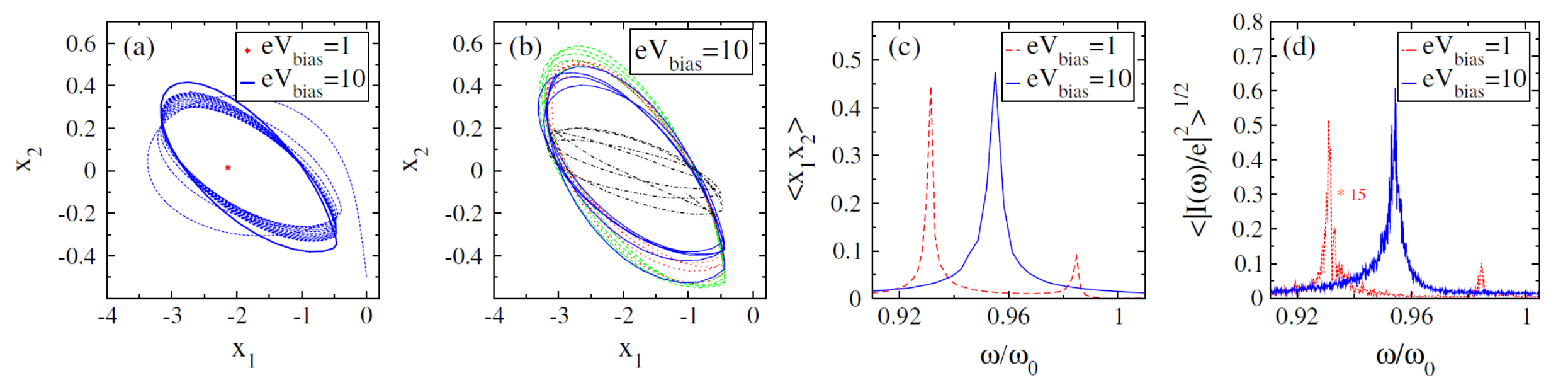}
	\end{center}
% \caption{Limit Cycles.}
\caption{ Results from a model-system consisting of two vibrational modes coupling to two electronic levels in the scattering region going beyond the harmonic approximation. 
(a) Limit cycle (blue solid line) and its approach (blue dotted line) at large bias vs stable oscillations at low bias
(red asterisk) from the Langevin dynamics without fluctuating force. (b) Several periods of typical trajectories with fluctuating forces for
the parameters of the limit cycle in (a). (c) Fourier transform of the correlation function $\left< X_1(t) X_2(t+\tau) \right>$. The limit cycle is signaled by
a single peak, as opposed to two peaks in the absence of a limit cycle. (d) The same signature appears in the current-current correlation
function, making the onset of limit-cycle dynamics in principle observable in experiment. 
Figure adopted from Ref. \cite{bode_scattering_2011} with permission. 
}
\label{fig:oppen11}
\end{figure}

In Ref.~\cite{bode_scattering_2011}, Bode and co-authors considered anharmonic
atomic dynamics driven by an electrical current using an equation similar to Eq.~(\ref{eq:lang1}).
There, the anharmonicity comes from the adiabatic $F_e$, which depends on the
atomic displacement $X$. They found that, beyond the harmonic instability, the
anharmonic forces stabilize the dynamics and the system goes into some kind of
limit cycles. The appearance of the limit cycles can be seen from the two-peak
to one-peak transition in the current and displacement correlation functions
(Fig.~\ref{fig:oppen11} (c), (d)).

\begin{figure}[!ht]
	\begin{center}
	\includegraphics[scale=0.4]{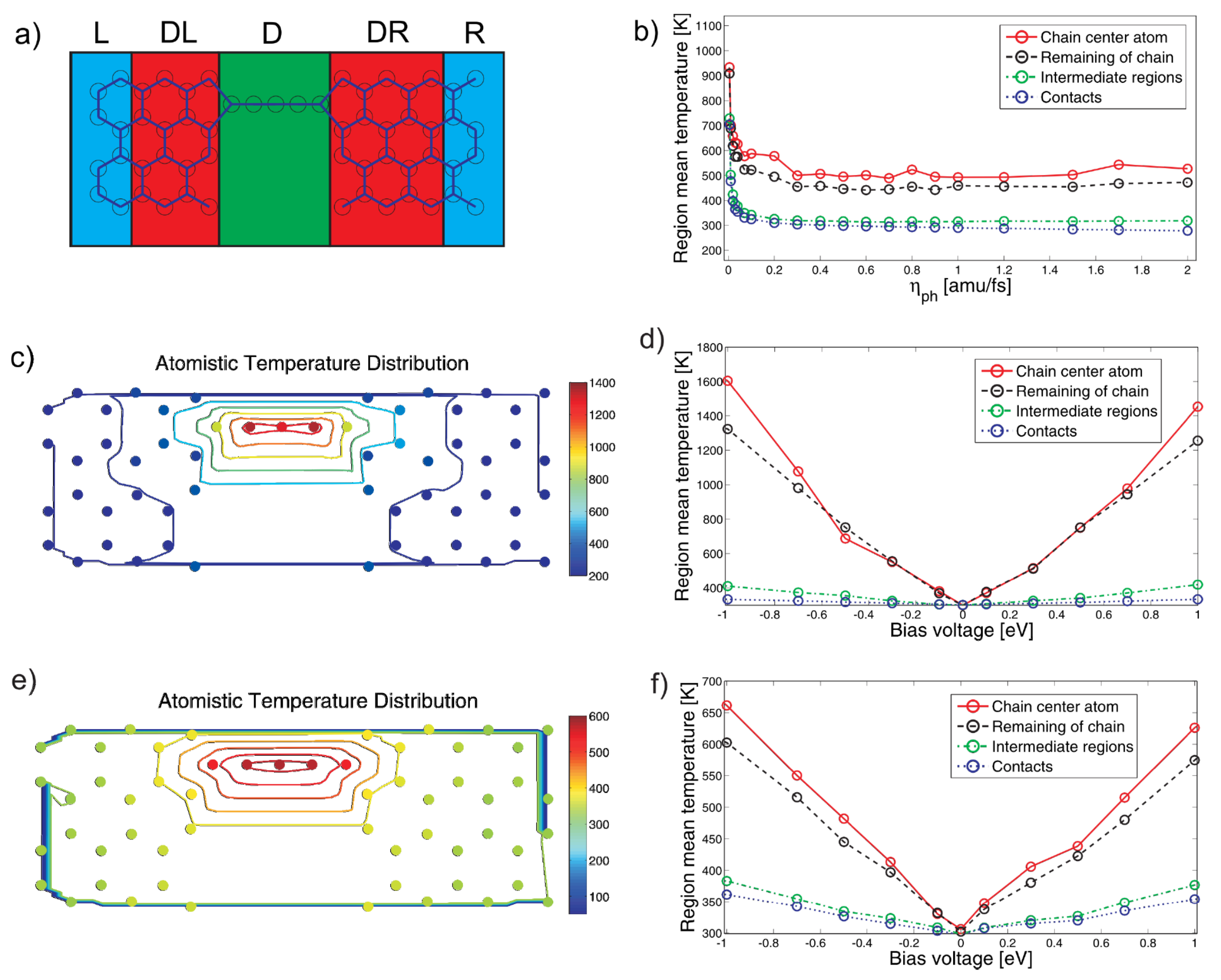}
	\end{center}
% \caption{MD simulation on the fly.}
\caption{(a) The system considered in the calculation, with single atomic carbon chain between
graphene nanoribbon electrodes.
(b) Temperature obtained from the atomic kinetic energy as a function of phonon friction.
(c-d) Obtained temperatures deduced from the atomic kinetic energy of different atoms within the harmonic approximation. 
(c) The simulations were run at $T = 300$ K and at $eV_b$ = 1 eV. 
(d)  The atomic temperature as a function of bias at $T=300$ K.
(e, f) Corresponding atomistic temperature distributions including 
the anharmonic interactions.  The anharmonic interactions redistribute 
part of the energy from the modes  in the chain to the bulk modes in the lead.
Figure adopted from Ref. \cite{lu_current-induced_2011} with permission. 
}
\label{fig:tuemd}
\end{figure}

The SGLE has also been used to study current-induced dynamics going towards a more realistic atomistic description of the
system. The authors have considered the dynamics of an carbon atomic chain between two
graphene nanoribbon electrodes\cite{lu_current-induced_2011,tue-thesis,christensen_current-induced_2016,gunst_phonon_2013}. A tight-binding model for the electronic 
structure and the Brenner potential for the inter-atomic interaction was used. 
The electronic structure was updated on
the fly during the MD simulation (Fig.~\ref{fig:tuemd}).  This kind of calculation
makes it possible to study current-induced dynamics using MD 
simulation. So far, the dynamics was based on empirical potentials. However, it would be
highly desirable to perform {\it at initio} MD based on f.ex. DFT electronic structure calculations. In order to achieve this goal, an efficient way
of performing the DFT calculations is crucial possibly including partly the non-equilibrium forces using DFT-NEGF\cite{brandbyge_origin_2003}. 

%%%%%%%%%%%%%%%%%%%%%%%%%%%%%%%%%%%%%%%%%%%%%%
%%%%%%%%%%%%%%%%%%%%%%%%%%%%%%%%%%%%%%%%%%%%%%
%\subsection{Molecular Dynamics simulations}
%%%%%%%%%%%%%%%%%%%%%%%%%%%%%%%%%%%%%%%%%%%%%%
%%%%%%%%%%%%%%%%%%%%%%%%%%%%%%%%%%%%%%%%%%%%%%

%+++++++++++++++++++++++++++++++++++++++++++++++++++++++++++++
\subsection{Nuclear quantum effect in molecules and solids}
%++++++++++++++++++++++++++++++++++++++++++++++++++++++++++++

\begin{figure}[!ht]
	\begin{center}
	\includegraphics[scale=0.4]{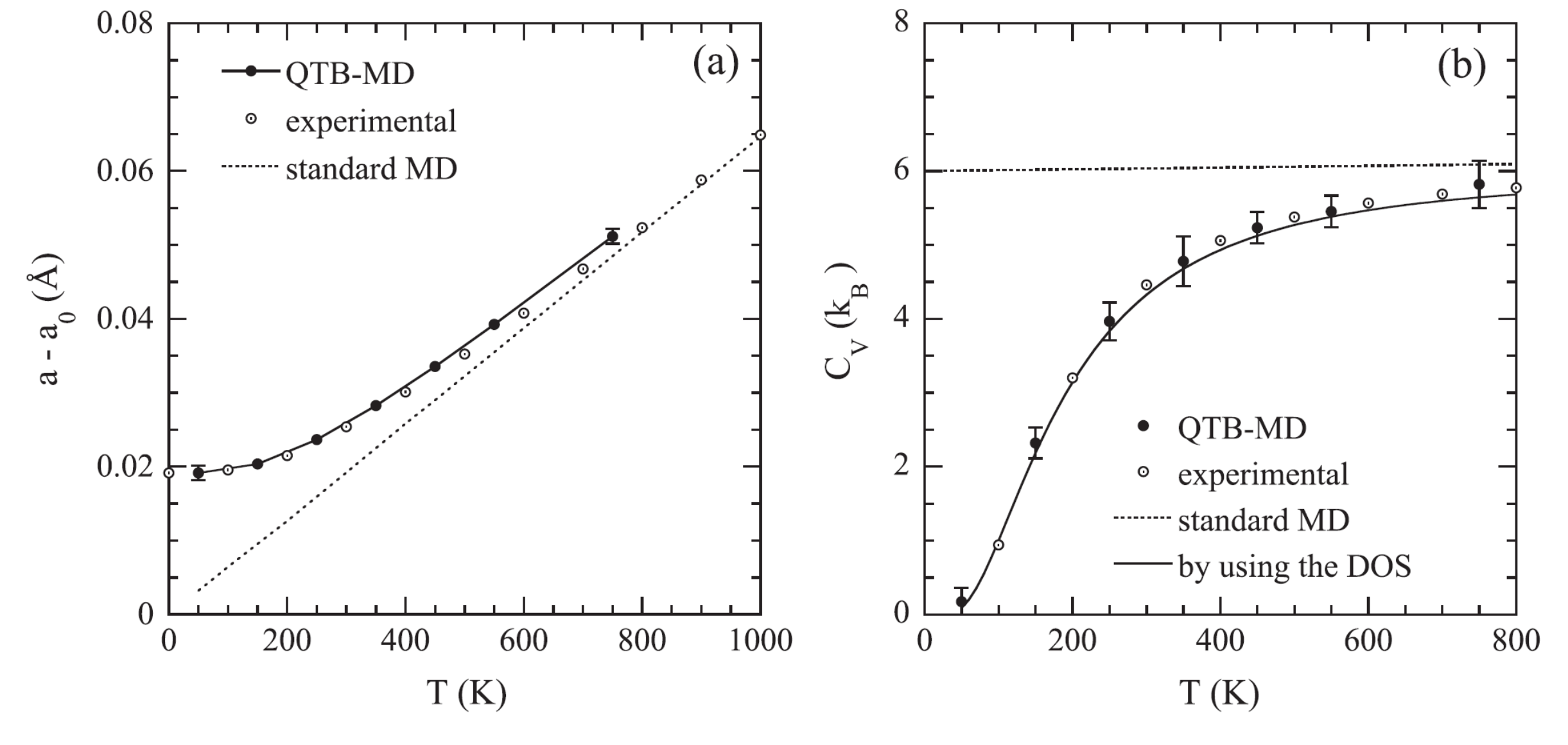}
	\end{center}
% \caption{Dammak, PRL 09 figure 3}
\caption{MgO crystal parameters predicted from MD with QTB (QTB-MD). 
(a) Temperature dependence of the lattice parameter $a$. The $a_0$ value is obtained by
extrapolating, to 0 K, the linear behavior observed at high temperature. 
The QTB-MD reproduces the experimental data at low temperatures. 
(b) Temperature dependence of the heat capacity per molecule $C_V$. 
The QTB values (obtained by differentiation of the
mean energy) agree with the experimental data and the results derived using the harmonic density of vibrational states (DOS). The
standard MD simulation gives reliable values only at temperatures higher than the Debye one (940 K).  Figure adopted from Ref. \cite{dammak_quantum_2009} with permission. 
}
\label{fig:dammak09}
\end{figure}

The nuclear quantum effect (NQE) has important implications for the physical and
chemical properties of molecules and crystals made from light elements and
stiff chemical bonds\cite{cazorla_simulation_2017}. One notable example is the
complicated behavior of the hydrogen bond under different conditions\cite{li_quantum_2011}.
To study the NQE, the dynamics of nuclei has to be treated quantum mechanically, and to this end  path-integral MD (PIMD) and Monte Carlo methods are normally used. 
%A semi-classical approach has its advantages in computational cost and physical transparency\cite{}. 
Recently, the semi-classical
Langevin equation was used to partially account for the nuclear quantum
effect\cite{ceriotti_nuclear_2009,dammak_quantum_2009,dammak_isotope_2012,ceriotti_colored-noise_2010,brieuc_zero-point_2016,brieuc_quantum_2016,ganeshan_simulation_2013,bronstein_quantum_2016,bronstein_quantum-driven_2014}.
The idea is to simply attach one \emph{artificial} quantum thermal bath (QTB) to each nuclear DoF.
For each element $X_i$, we have
\begin{equation}
	\ddot{X}_i - F_{I,i} = -\gamma_i \dot{X}_i + \chi_i.
	\label{eq:qtb}
\end{equation}
\revision{Equation~(\ref{eq:qtb}) has similar form as Eq.~(\ref{eq:lang1}). 
Its left side represents the original equation of motion. Different from Eq.~(\ref{eq:lang1})
where the system couples to an electron bath, 
in Eq.~(\ref{eq:qtb}) $X_i$ couples to an \emph{artificial} QTB.}
The friction coefficient, $\gamma_i$, is an adjustable parameter, and the correlation
function of the fluctuations fulfills the fluctuation-dissipation relation:
$\langle \chi_i(\omega) \chi_i^*(\omega)\rangle = \gamma_i \omega
\coth\left(\omega/2k_BT)\right)$.  This is a colored noise spectrum
that depends on $\omega$. In the high temperature limit, it reduces to the
well-known relation: $\langle \chi_i(\omega) \chi^*_i(\omega)\rangle = 2\gamma_i
k_B T$, where $\omega$-dependence can be ignored and the noise spectrum becomes
white.

%(3) It is a \emph{first-principles} method, where the microscopic interactions can
%be taken into account in principle, although certain approximation are necessary in practice.
The advantages of this approach is that it introduces almost no extra 
computational cost compared to standard classical MD.
Its performance in determining many physical
properties has been calibrated through comparison to the fully quantum
mechanical approaches\cite{ceriotti_nuclear_2009,dammak_quantum_2009,brieuc_zero-point_2016,ganeshan_simulation_2013,bronstein_quantum_2016,bronstein_quantum-driven_2014,barrozo_comment_2011,dammak_dammak_2011}.

Dammak \emph{et al.} showed that temperature dependence of MgO's lattice
constant and its heat capacity calculated from the SGLE agree well with
experimental measurements\cite{dammak_quantum_2009}(Fig.~\ref{fig:dammak09}). 
Bronstein and coauthors studied the NQE on the phase transition
in pure and salty ice at high pressure\cite{bronstein_quantum-driven_2014,bronstein_quantum_2016}. They found excellent agreement between the 
experimental and the theoretical results using the QTB model.
Further applications include vibrational properties of polyatomic molecules
\cite{calvo_vibrational_2012,calvo_atomistic_2014}, 
isotope effect\cite{dammak_isotope_2012},
shocked-compressed molecules\cite{qi_simulations_2012,shen_quantum_2016},
spin-phonon dynamics\cite{woo_quantum_2015,swinburne_low_2017,bergqvist_realistic_2017}.

Ceriotti \emph{et al}. took an important step further\cite{ceriotti_nuclear_2009,ceriotti_langevin_2009,ceriotti_colored-noise_2010,ceriotti_efficient_2010,ceriotti_nuclear_2016,ganeshan_simulation_2013,brieuc_quantum_2016}.  We have mentioned that, given the full
Hamiltonian of the global system, the parameters entering the SGLE are derivable from
the microscopic Hamiltonian. Ceriotti and coauthors have performed a `reverse-engineering'
of the QTB, making full use of the flexibility
in choosing the bath parameters for target applications. This concept has been termed
`Langevin thermostat \emph{\`a la carte}'.
They showed that through systematic optimization, the properties of the QTB can be tailored
to display desired sampling features, i. e., selective coupling of the thermal bath to 
target certain vibrational modes. When combined, the QTB
can significantly improve the convergence and scalability of PIMD to reach a performance comparable to that of the standard Nos\'e-Hoover chain thermostat.

As a semi-classical approach, the SGLE can only account partially for the quantum
effects. More importantly, it has the problem of zero point energy (ZPE)
leakage \cite{bedoya-martinez_computation_2014}. This problem arise from the fact that in the SGLE, the ZPE
is stored in each harmonic mode classically. 
In the presence of anharmonic
couplings, even at zero temperature, the classically stored ZPE can redistribute 
between different vibrational modes. This unphysical energy flow from the high frequency modes to the low
frequency vibrational modes results in a wrong energy distribution (ZPE leakage).
The average energy of each vibrational mode is determined by its coupling to
the thermal bath characterized by $\gamma$ in the Langevin equation and to
other modes determined by the anharmonic coupling. The problem
is common to methods based on classical trajectories. Different solutions of
this problem have been
proposed\cite{bedoya-martinez_computation_2014,czako_practical_2010,brieuc_zero-point_2016}.
Brieuc \emph{et al}. have studied this effect carefully in both model and realistic
structures\cite{brieuc_zero-point_2016} and found that, in most cases, by
simply increasing $\gamma$, the energy exchange between the thermal bath and the vibrational mode
becomes dominant. In this way the ZPE leakage becomes relatively small. Taking notice of this problem, except for very anharmonic systems, the QTB can then give reasonable results for many properties. 

%+++++++++++++++++++++++++++++++++++++++++++++++++++++++++++++
\subsection{Quantum thermal transport in nanostructures}
\begin{figure}[!ht]
	\begin{center}
	\includegraphics[scale=0.8]{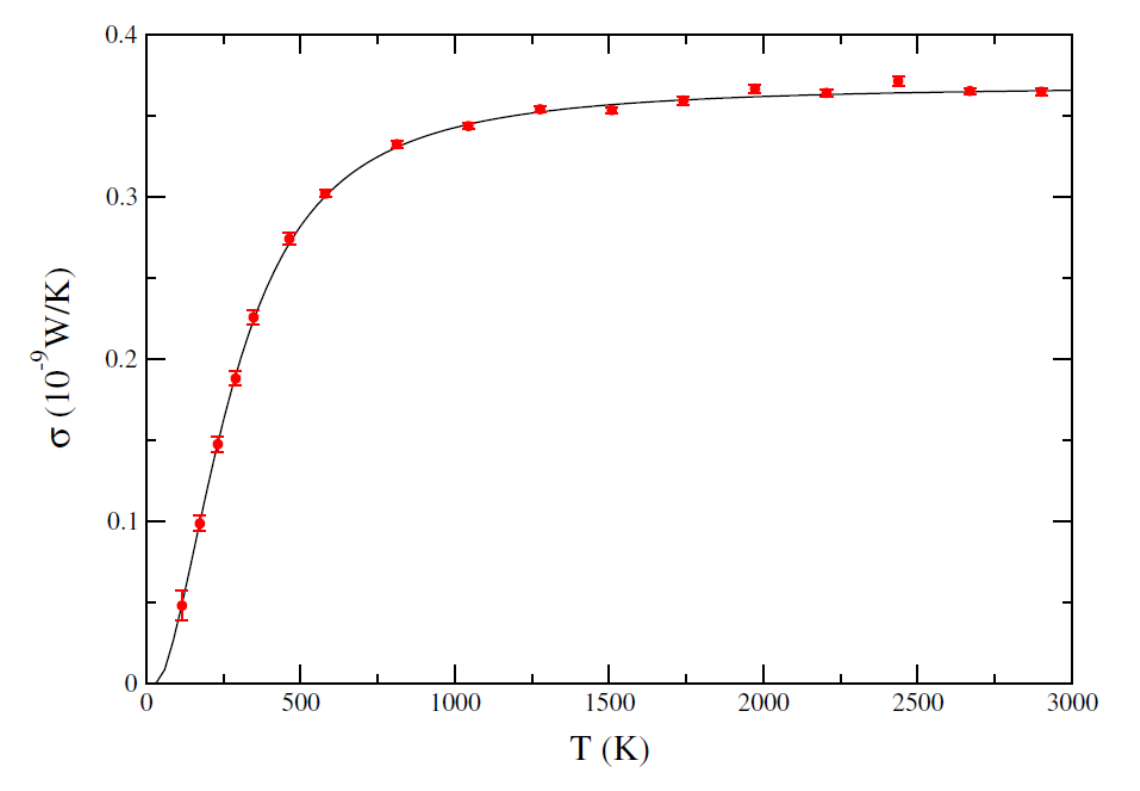}
	\end{center}
% \caption{Jian-Sheng Wang, PRL 07 figure 1}
\caption{ Thermal conductance $\sigma$ as a function of
temperature for a 1D harmonic chain with on-site potential. 
The solid line is the NEGF result, while
the symbols are MD results. They agree with each other.
Figure adopted from Ref. \cite{wang_quantum_2007} with permission. 
}
\label{fig:wang07}
\end{figure}

%++++++++++++++++++++++++++++++++++++++++++++++++++++++++++++++
By attaching the system to multiple QTBs, one may study
phonon heat transport using the SGLE\cite{dhar_heat_2001,segal_thermal_2003,wang_quantum_2007,wang_quantum_2008,roy_crossover_2008,kantorovich_generalized_2008-1,kantorovich_generalized_2008,lu_coupled_2009,li_colloquium_2012,kosevich_effects_2013}.
For linear harmonic systems, the ZPE leakage is not present, and
has been shown analytically that, the SGLE produces exact results
consistent with fully quantum mechanical approaches
(Fig.~\ref{fig:wang07})\cite{dhar_quantum_2003,dhar_heat_2006,wang_quantum_2007}.
Figure~\ref{fig:wang07} compares numerically the thermal conductance of a model
harmonic chain calculated from the SGLE and the NEGF method.  The agreement is
due to the quantum statistics of the mode occupations in the thermal bath. For linear harmonic systems this can efficiently propagate into the system through the system-bath
coupling. This is especially important for systems with light atoms whose
Debye frequency is high, and classical statistics fails. For example, the room temperature
thermal conductivity of carbon materials can not be predicted directly from
classical MD due to the large Debye frequency. Similarly, Langevin equations for
electrons are also derived and shown to be consistent with the standard
Landauer or NEGF result for noninteracting
electrons\cite{dhar_quantum_2003,lu_coupled_2009}.

\begin{figure}[!ht]
	\begin{center}
	\includegraphics[scale=0.4]{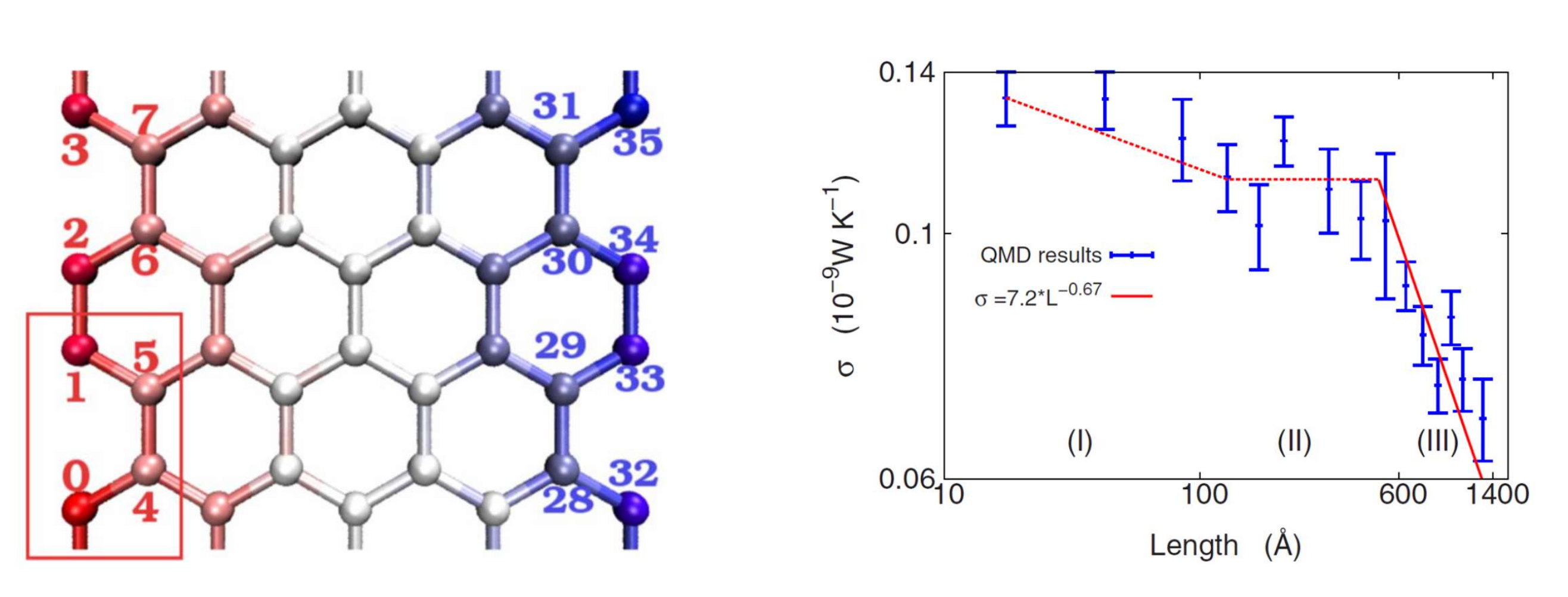}
	\end{center}
% \caption{Jian-Sheng Wang, PRB 09 figure 6}
\caption{Left panel: The structure for an armchair graphene
strip. The box (red) is the unit cell. The length of the ribbon along transport direction (horizontal direction) is varied while keeping the average temperature at $300$ K, while periodic boundary condition is applied to the perpendicular direction. 
Right panel: The dependence of the thermal conductance
on the length of the system at 300 K in logarithmic
scale. Phonon transport changes gradually from ballistic to diffusive
with increasing length of the system.
Figure adopted from Ref. \cite{wang_molecular_2009} with permission.
}
\label{fig:wang09}
\end{figure}

On the other hand, including the anharmonic interactions, at high temperatures
where the quantum mechanical effect is not important, the SGLE yields results
consistent with classical MD simulation\cite{wang_quantum_2007}.  Thus,
the transition of heat transport from quantum, ballistic to the classical, diffusive
regime can be studied using this
approach\cite{wang_quantum_2007,wang_molecular_2009,saaskilahti_thermal_2013,roy_crossover_2008}.  Figure~\ref{fig:wang09}
shows the thermal conductance ($\sigma$) of a graphene nanoribbon as a function
of ribbon length. When the ribbon is short, $\sigma$ does not depend on the
length, the system is in the ballistic transport regime. When the length is $>
600$ nm, $\sigma$ decreases with length as $\sim
L^{-0.67}$\cite{wang_molecular_2009}.

Intuitively, the ZPE leakage is not expected to have a large
influence on the real space heat transport. The modes contributing most to heat
transport should be traveling waves that are delocalized in real space.  Thus,
the energy leakage between different modes does not take place locally, while
the heat current is calculated locally. But a case study shows the opposite.
A comparison of classical, semi-classical
and experimental results of thermal conductivity of solid argon shows that,
the SGLE approach behaves much worse than the classical MD
\cite{bedoya-martinez_computation_2014}. Presumably,
the failure of the SGLE is due to the ZPE leakage, while the 
accidental cancellation of the two errors makes the classical MD results 
agree better with the experiments. This clearly shows the limitation of the 
SGLE, but more careful studies are needed before one can draw any conclusions
on the effect of ZPE leakage on the thermal conductivity 
predicted from the SGLE.

One important advantage of the GLE approach is that it is a \emph{first
principles} approach, in the sense that the parameters entering the SGLE
can be calculated from the microscopic Hamiltonian. We can calculate the friction matrix and fluctuating force correlations based on this, as long as the
bath and the system-bath coupling are linear. Even in the classical limit, the
GLE is an ideal method to perform MD simulation, due to its stronger
theoretical foundation compared to other approaches. For example, it fulfills
the fluctuation-dissipation relation and ensures that the system can reach the
canonical distribution in the long time
limit\cite{kantorovich_generalized_2008}. Recently, algorithms aiming
at application to realistic materials have been developed to efficiently treat
the memory effect in the friction kernel and the colored noise
spectrum\cite{stella_generalized_2014,ness_applications_2015}.  
This opens the possibility of its application to realistic materials.

%\subsection{Radiative heat transfer}
%\cite{chalopin_radiative_2011}

%%%%%%%%%%%%%%%%%%%%%%%%%%%%%%%%%
\section{Conclusions}
%%%%%%%%%%%%%%%%%%%%%%%%%%%%%%%%%
In this review, based on the influence functional approach, 
the classical Langevin equation is extended to include 
the quantum statistics and nonequilibrium features of the reservoir degrees of freedom. 
We have considered both phonon and electron reservoirs. 
These extensions result in a semi-classical generalized Langevin equation (SGLE),
which can be used to study different problems that are difficult
to handle using classical MD.  
The nuclear quantum  effect in materials can be partly 
included through coupling to the quantum phonon baths. Phonon thermal transport can be studied by coupling the system to several phonon reservoirs with different temperatures. 
Atomic vibration, translation and rotation of adsorbates on 
metal surfaces are damped through the electron-hole pair excitation in the metal electrons. 
This is taken into account as the electronic friction in the SGLE.
For nano-scale conductors in the presence of a current flow in the electronic reservoir, 
several interesting effects are predicted from the SGLE.
Apart from the electronic and phononic reservoirs, a quantum thermal
bath representing black body radiation has recently been used to study the radiative heat transfer from a black body to nearby dielectric 
nanoparticles\cite{chalopin_radiative_2011}. This is possible since
eigenmodes of electromagnetic waves and phonons are both 
represented by a set of harmonic oscillators. 
These results greatly extend the range of applications of the MD 
methods.  

We also discussed the technical problems and available solutions 
in order to use the method to study realistic systems. 
The non-Markovian friction kernel can be transformed to a Markovian
one by introducing auxiliary degrees of freedom.
The colored noise can be generated on the fly during the MD 
simulation. 
The implementation of the quantum phonon bath in available molecular
dynamics codes will further accelerate its application to the problems mentioned above. Several groups are working on this now, and it is believed that, these developments will enrich and widen the applications 
of MD.

\section*{Acknowledgements}
The authors would like to thank Jian-Sheng Wang, Baowen Li, Tchavdar Todorov, 
Daniel Dundas, Nuo Yang, Tue Gunst, Rasmus B. Christensen, Nick R. Papior  for
discussions and collaborations on this subject. Financial support from the
National Natural Science Foundation of China (grant number: 61371015), and
the Villum Foundation (to MB, VKR00013340),
is
gratefully acknowledged.

%%%%%%%%%%%%%%%%%%%%%%%%%%%%%%%%%
\section*{References}
\bibliography{extra,gle-review}
%%%%%%%%%%%%%%%%%%%%%%%%%%%%%%%%%

\end{document}